\documentclass[12pt]{article}
\usepackage{epsfig, amssymb}
\usepackage{amsmath}
\usepackage{physics}
\usepackage{graphicx,epsfig}
\usepackage{color} 
\usepackage{cite}
\usepackage[title]{appendix}
\usepackage{hyperref}
\usepackage{tikz}
\usepackage{mathtools}
\usepackage{tikz}
\usetikzlibrary{decorations.markings,arrows.meta}
\usepackage{pgfplots}
\usepackage[normalem]{ulem} 
\pgfplotsset{compat=1.18}
\begin{document}
\thispagestyle{empty}
\begin{flushright}
\end{flushright}

\bigskip

\begin{center}
{\Large \textbf
{Stochastic dynamics from O(2N) fractional Laplacian vector model and O(N) vector model free energy at finite temperature}} \\ 
\vspace{2cm} WooCheol Shin${}^{a}$\footnote{e-mail:syc5216@hanyang.ac.kr}, Jun Hyuk Lee${}^{a}$\footnote{e-mail:junseyu2@gmail.com}, {Ji-seong Chae${}^{a}$\footnote{e-mail:jiseongchae17@gmail.com} and Jae-Hyuk Oh${}^{a}$\footnote{e-mail:jaehyukoh@hanyang.ac.kr}
}

\vspace{1cm}
  {\it
Department of Physics, Hanyang University, Seoul 04763, Korea${}^{a}$\\
 }
\end{center}

\vspace{0.3cm}
\begin{abstract}
We explore O(2N) vector model with fractional Laplacian, $\sqrt{-\nabla^2}$ in $d$-dimension
and its Hamiltonian dynamics which is described by a Schrodinger type equation. 
This equation is a kind of current conservation equation, where one can define a current $j(\phi^a)$ of a probability $P(\phi^a)$, where $\phi^a$ is the O(2N) vector field. Naturally, Gibbs entropy $S=-\int [D\phi^a] P(\phi^a)\log P(\phi^a)$ can be considered to explore the system. We realize that this Gibbs entropy of the  O(2N) vector model with fractional Laplacian is matched with free energy of O(N) vector model at finite temperature, $1/\beta$ with a deformation, $\mu$ in $d$-dimension. The precise map between the stochastic fictitious time $t$ and the inverse temperature $\beta$ is $\beta=2t$. Therefore, the temperature dependence of the thermal O(N) vector model can be realized as the dynamics of a time dependent solution satisfying Schrodinger type equation. This free energy is obtained by putting O(N) vector model in $S^1\times \mathbb R_d$, where $S^1$ is thermal circle with its periodicity $\beta$. To get $d$-dimensional theory, we sum up all possible frequencies on the circle(so called Matsubara frequency summation) which gives $d$-dimensional thermal partition function. We note that the nontrivial $t$-dependence appears beyond classical limit. 
To take into account quantum effects, we solve the Hamiltonian dynamics by keeping $\hbar$ corrections. The spectral deformation is mediated by a parameter $l$ such that $\mu=\beta^{-1}\log l$ and so we call this $l$-deformation. This is related to the initial boundary condition of the Schrodinger equation. We also note that the two theories are not equivalent to each other and we just check their correspondence at the level of one-loop determinant, i.e. zero point function in the note.
\end{abstract}
\newpage
\tableofcontents

\section{Introduction}
O(N) vector model is widely studied in various contexts, such as condensed matter theories \cite{4,Moshe:2003xn}, string theory via holography\cite{Klebanov:2002ja,Giombi:2009wh,Giombi:2013fka,Giombi:2016ejx,Maldacena:2011jn} and so on. 
In condensed matter theory, the O(N) vector model is continuous version of the discrete N-dimensional spin system. Therefore it is natural that the theory has a constraint, namely $\sum_{a=1}^N \phi^a \phi^a=N$, where $\phi^a$ is N-collection of scalar fields which is in fundamental representation of O(N) rotation generators. This constraint ensures that the magnitude of the O(N) vector field is fixed. Such  a constraint is realized in the form of $\delta$-function insertion into partition function in the path integral formulation of the theory.

A noticeable methodology to deal with this constraint is large-N mean field theory. The $\delta$-function constraint is exponentiated by employing an auxiliary field. It is well known that in the limit that $N\rightarrow\infty$, saddle point approximation becomes accurate and the variation of the action in the exponent of integrand of the partition function with respect to the auxiliary field provides exact equation of the system. This is so-called gap equation. The gap equation may or may not give mass gap to the spectrum of the theory, which depends on the strength of the coupling in the theory.

On the other hand, O(N) vector model in holographic context is somewhat different, The theory in this context is O(N) vector model with $(\sum_{a=1}^N \phi^a \phi^a)^2$ interaction. In holography, the theories at the conformal fixed points are extensively studied. There are two kinds of fixed point: one is the free theory limit and another is the theory at the critical coupling, a kind of Wilson-Fisher fixed point in Infra-red region. The spectrum is gapless since they are conformal field theory. We note that for the O(N) vector model in holographic context, there is no constraint such as $\sum_{a=1}^N \phi^a \phi^a=N$. So the field theory is just collection of non-interacting N free scalars in such a limit that the self-interaction vanishes.
Another interesting part is that in such a fixed point there emerges a new type symmetry, so called higher spin symmetry. In the free theory limit, the symmetry is exact but in the critical coupling limit, it is violated with $\mathcal O$(1/N)-terms and so becomes exact in the limit of $N\rightarrow \infty$.

In this note, we suggest a new frame to understand O(N) vector model at finite temperature,
stochastic dynamics of the vector model with fractional Laplacian. More precisely, we suggest a way to understand O(N) vector model at finite temperature with a (spectral) deformation via the dynamical framework: stochastic dynamics at the zero mass gap sector.


The motivation why we consider this is two folds. First motivation is to provide more concrete and illustrative relationship between vector models with standard Laplacian and fractional Laplacian theories. There appears a fractional Laplacian theory as a generating functional of dual boundary field theory from the gravity theory with standard kinetic terms in holography, for example scalar field theory in AdS space and its dual field theory generating functional. In scalar field model in asymptotically AdS space, the near boundary expansion is given by
\begin{equation}
\phi(r,x)=\phi(x)r^{\frac{d}{2}-\nu}+a(x)r^{\frac{d}{2}+\nu},
\end{equation}
where $x$ is the boundary directional coordinates, $r$ is AdS radial coordinate and $\nu=\sqrt{\frac{d^2}{4}+m^2}$. $m$ is the mass of the scalar field and the boundary is at $r=0$. $\phi$ is the coefficient of non-normalizable mode which corresponds to source in the generating functional $W(\phi)$ of the boundary field theory and $a$ is the coefficient of the normalizable mode which does to a composite operator of the boundary field theory. The boundary generating functional in momentum space is given by
\begin{equation}
W(\phi)=\int d^dk \phi(k)\phi(-k)|k|^{2\nu},
\end{equation}
and the factor of $|k|^{2\nu}$ corresponds to $(-\nabla^2)^{\nu}$ operator in position space. 
This does not mean that the corresponding field theory is fractional Laplacian theory but its effective theory 
probably does. Above its unitarity bound in the dual field theory and Breitenlohner–Freedman bound in the bulk theory, the effective theory is probably well defined. 

Once the boundary value of the scalar field, $\phi$ is held fixed, then that corresponds to standard quantization. However, as long as the mass of the scalar field stays in the unitarity window, $-\frac{d^2}{4} \leq m^2 \leq -\frac{d^2}{4}+1$ also known as $0\leq \nu \leq 1$, alternative quantization is also possible. In this context, the role of $\phi$ and $a$ switched.
%
%
One of the good example as such is conformally coupled scalar theory in Euclidean AdS space. It mass is $m^2=-\frac{d^2-1}{4}$ and so $\nu=1/2$.

The relation between holographic gravity model and its boundary generating functional is mathematically the same with another interesting dynamical framework, which is stochastic dynamics\cite{Damgaard:1987rr,Dijkgraaf:2009gr,Mansi:2009mz}. Especially, the authors in\cite{Oh:2012bx,Jatkar:2013uga,Oh:2013tsa,Oh:2015xva,Oh:2021bxx,Kim:2023bhp,Lee:2023ynb,Chae:2024zyn} reformulate holographic Wilsonian renormalization group by employing Langevin dynamics, equivalently Fokker-Planck dynamics. 

Stochastic dynamics is basically dynamics of statistical system contacting with thermal bath, which is a relaxation process from non-equilibrium state to equilibrium one. As we will review in Sec.\ref{A brief review of stochastic quantization method}, stochastic dynamics is described by Langevin equation, which is a kind of diffusion equation. This equation describes evolution of the statistical system from non-equilibrium to equilibrium by employing fictitious time $t$. As $t\rightarrow \infty$, the system settles down an equilibrium state, In the equilibrium, the stochastic partition function can be identified to Euclidean partition function with a certain theory, $S_E$, which is called Euclidean action, once $k_BT=\hbar$, where $T$ is the temperature of the thermal bath and $k_B$ is Boltzmann constant. This fictitious time, $t$ evolution of the system can be also described by Fokker-Planck equation, which is a Schrodinger type equation with Hamiltonian, which is call Fokker-Planck Hamiltonian made out of $S_E$. The main claim in \cite{Oh:2012bx,Jatkar:2013uga,Oh:2013tsa,Oh:2015xva,Oh:2021bxx,Kim:2023bhp,Lee:2023ynb,Chae:2024zyn} is that once one identifies $S_E(\phi)=-2W(\phi)$ and $t=r$, the holographic Wilsonian renormalization group equation coincides with Fokker-Planck Hamiltonian dynamics. 

The scalar theory can be easily generalized to free O(N) vector model and we suspect if the vector models with standard Laplacian kinetic terms and fractional Laplacian ones are related each other. Therefore, we will demonstrate a concrete example in this note.

The second motivation is if the O(N) vector model can reformulate by employing a more dynamical framework. The large-N mean field theory of O(N) vector model is successful since the mean field theory can explain macroscopic properties of the system quite well. In the Large-N limit, solving gap equation is to find saddle points of the effective theory. On the other hand, stochastic dynamics describes relaxation process to stationary states through dynamical evolution.  This suggests that the determination of the expectation value of mean field may be reformulated as a dynamical evolution problem rather than a purely static saddle point condition. Therefore, finding saddle points of the system might be translated into solving dynamical differential equation in stochastic frame.



What we find is that stochastic dynamics of fractional Laplacian($\sqrt{-\nabla^2}$) theory can describe dynamical evolution of the O(N) vector model at finite temperature. In O(N) vector model side, this describes evolution from one local equilibrium with $\beta_1$ to another equilibrium with $\beta_2$, where $\beta_i$ is the inverse temperature of the system. 
The fractional Laplacian theory is composed of 2N-real scalars without self-interaction and so it enjoys O(2N) symmetry.

We compare the two distinct theories:
\begin{itemize}
\item One is mean field theory limit of O(N) vector model on $S^1\times \mathbb R_d$, where $S^1$ is thermal circle with periodicity $\beta$ without mass gap, so it becomes vector model at finite temperature. The theory is gapless in a way that either we do not impose the constraint, $\sum_{a=1}^N \phi^a \phi^a=1$ or we impose the constraint but in the small coupling limit without the spectral deformation, the theory has zero mass gap.

\item Another is O(2N) real scalar model on $\mathbb  R_+\times \mathbb R_d$ without self-interaction and its kinetic term is composed of fractional Laplacian, $\sqrt{-\nabla^2}$. The $\mathbb R_+$ is a half of the real line and  indicates the stochastic time direction for the Fokker-Planck dynamics.
\end{itemize}

We realize that the two theories provide the same (renormalized) one loop determinant once we identify the quantities appearing in the both sides as follows:
\begin{itemize}
\item $\beta=2t$, where $\beta$ is the inverse temperature for the $O(N)$ vector model, i.e. the periodicity of the thermal circle $S^1$ and $t$ is the stochastic time, which lives on $\mathbb R_+$.
\item $l=l^{(H)}$. The $l$ is a parameter for a spectral deformation for O(N) vector model, where the spectral deformation $\mu=\beta^{-1}{\log l}$. $l^{(H)}$ is a parameter for imposing initial boundary condition of stochastic dynamics.
\end{itemize}
The normalized one loop determinants in the both sides are obtained as follows respectively. In the O(N) vector model, we compute free energy density, $\mathcal F(\beta, l)=F(\beta, l)/V_d$, which is basically logarithm of its partition function, where $V_d$ is d-dimensional spatial volume. The free energy density is not regular at $\beta=\infty$, and so we regularize the free energy density by subtracting $\mathcal F(\infty,l)$ from it.

On the other hand, the Fokker-Planck dynamics from O(2N) fractional Laplacian vector model provides its probability distribution, $P(\phi,t)$ as a function of $t$. The $P(\phi,t)$ is a non-trivial weight at a given stochastic time $t$ for the partition function $Z(t)$. The $P(\phi,t)$ is defined as $Z(t)=\int [\mathcal D \phi]P(\phi,t) $. We note that the partition function is already normalized so the probability has a form of $P(\phi,t)=\exp[-\int d^dkdt K(t)\phi^a(k,t)\phi^a(-k,t)]/P_0$, where $K(t)$ is Kernel for the free O(2N) vector model and $P_0$ is the normalization factor. Extracting one loop determinant of the kernel $K(t)$ from it is to compute Gibbs entropy density of the probability, $s(t)=-(V_d)^{-1}\int [\mathcal D \phi]P(\phi,t)\log P(\phi,t)$. 
Once we impose the relations that we mentioned above as $\beta=2t$ and $l=l^{(H)}$, we can observe the following remarkable coincidence as
\begin{itemize}
\item $\beta [\mathcal F(\beta,l)-\mathcal F(\infty,l)]=s(t,l^{(H)})-s(\infty,l^{(H)})$ ,where we identify both $\beta=2t$ and $l=l^{(H)}$.
\end{itemize}

This result implies that the temperature dependence of O(N) vector model can be interpreted as a dynamical relaxation process satisfying probability conservation equation with fractional O(2N) scalar model.

One possible interpretation for the result is as follows. Suppose there is O(N) vector field system which contacts with thermal reservoir with its temperature zero i.e $\beta=\infty$. The O(N) vector system is much smaller than its reservoir. Let us take the scale of the system is $L$. For every intermediate fictitious time $t$, the system settles down in local equilibrium instantaneously. This could be a point particle limit of the system. The evolution of the system is adiabatic process, where $L \ll t$. This means that the size of the system is much smaller than the stochastic time scale. Every intermediate fictitious time, $t$, one can define equilibrium of the system with given temperature, $\beta=2t$.



\section{A brief review of stochastic quantization method}
\label{A brief review of stochastic quantization method}

\paragraph{Brownian motion} Physical motivation of stochastic quantization originates from Brownian motion.
A motion of a particle {undergoes} incessant random collisions with its environment and it can be described by introducing a random noise.
When the surrounding is thermal and viscous, the motion is diffusive and the noise becomes thermal noise depending on its temperature.
The original Langevin equation describing Brownian motion for a non-relativistic particle moving in $d$-spatial dimension with time $t$ reads
\begin{equation}
m\ddot{\vec{x}}(t) = -\alpha \dot{\vec{x}}(t) + \vec{F}(\vec{x},t) \text{  where  } 
\vec{F}(\vec{x},t) = -\nabla V(\vec{x}) + \vec{\eta}(t),
\end{equation}
where $m$ is the particle mass, $V(\vec{x})$ is a background potential, $\alpha$ is so called friction constant and $\vec{\eta}(t)$ is the random force that the particle receives from its environment.
In the presence of a potential, its effect is to introduce an external force to the particle like a charged particle in external electric field.
However, we consider the special case $V(\vec{x})=0$ and then this external force vanishes and the equation reduces to free Brownian motion, corresponding to purely diffusive dynamics.
We assume that the random force $\vec{\eta}(t)$ is taken to be Gaussian white noise.
Its probability distribution is defined as
\begin{equation}
\label{Gaussian white noise distribution}
P[\vec{\eta}(t)]=  \frac{\exp\left[
-\frac{1}{4\lambda}\int_{t_0}^{t} dt' \vec{\eta}^2(t')
\right]}{ \int \mathcal D \vec{\eta}\exp\left[
-\frac{1}{4\lambda}\int_{t_0}^{t} dt'' \vec{\eta}^2(t'')
\right]}.
\end{equation}
Normalization condition,
\begin{equation}
\label{Gaussian white noise normalization}
1=\int \mathcal{D} \vec{\eta}  P[\vec{\eta}(t)].
\end{equation}
is satisfied.
The expectation value of a physical variable $g[\vec{\eta}(t),t]$ with respect to this probability is given by
\begin{equation}
\langle g(t) \rangle \equiv \int \mathcal{D} \vec{\eta}  P[{\vec{\eta}}] g[\vec{\eta}(t),t].
\end{equation}
For instance, the correlation functions between noise fields are 
\begin{align}
\label{noise-corealtion-basic}
& \langle \vec{\eta}(t) \rangle = 0, \\
& \langle \vec{\eta}_{i}(t_1)\vec{\eta}_{j}(t_2) \rangle = 2\lambda \delta_{ij}\delta(t_1-t_2).
\end{align}
This can be easily obtained by introducing a source $J(t)$ to the partition functional as
\begin{equation}
Z[\vec{J}]=\int_{-\infty}^{\infty} \mathcal{D} \vec{\eta} P[\vec{\eta}]\exp \left(\int dt \vec{J}(t) \cdot \vec{\eta}(t) \right).
\end{equation}
and setting $J=0$ after taking functional derivatives of the partition function with respect to $J(t)$.

A few remarks are in order.  First, we note that such a particle motion under the random noise is so called Markovian process. This means that the state of the particle depends only on the last collision with its environment but it does not depend on its states before the last collision. This Markov process implies rapid loss of memory of the particle states and so its final destination does not depend on the initial condition at all.
Second, one can compute correlation functions between particle velocities, such as $\langle \dot {\vec {x}} \cdot \dot {\vec {x}}\rangle$ by solving the Langevin equation. The solution is a form of $\dot{\vec {x}}(\vec{\eta}(t))$. By employing the noise correlation functions given in (\ref{noise-corealtion-basic}), one can compute the expectation value of the particle energy, which is given by
\begin{equation}
\langle E(t) \rangle=\frac{1}{2}m \langle \dot{ \vec{ x}}(t) \cdot \dot{ \vec{ x}}(t) \rangle=\frac{d \lambda}{2 \alpha}\left( 1- e^{-\frac{2\alpha}{m}t}\right)
\end{equation}
Using Einstein relation $\lambda =k_{B}T \alpha $, the energy expectation value approaches to $\frac{d}{2}K_BT$ as $t \rightarrow \infty$. This result is consistent with equipartition principle.
Finally, we note that we apply this stochastic quantization method to quantize Euclidean field theory. For this purpose, we switch $K_BT$ to $\hbar$ and Einstein relation is changed as $\lambda=\hbar \alpha$. In the following, we choose $\alpha=\frac{1}{2}$ and $\lambda=\frac{\hbar}{2}$.

\paragraph{Stochastic dynamics via Langevin equation}Stochastic quantization is a quantization method for Euclidean field theory by employing Langevin equation and white Gaussian noise, This stochastic quantization method is very much similar with the Brownian motion that we introduced in the last paragraph. To illustrate this, let us consider an Euclidean scalar field theory.
The scalar field $\phi(x)$ is defined in a $d$-dimensional Euclidean space and suppose we want to quantize its theory being given by classical action $S_{\mathrm{cl}}[\phi]$. Now, we take a promotion such that the scalar field is to become stochastic time(also known as fictitious time) ``$t$'' dependent as $\phi(x)\rightarrow\phi(x,t)$, The form of Langevin equation is given by
\begin{equation}
\label{stochastic field with langevin eq}
\frac{\partial \phi(x,t)}{\partial t}
= - \frac{1}{2} \frac{\delta S_{cl}[\phi;t]}{\delta \phi(x,t)}
+ \eta(x,t),
\end{equation}
where $\frac{1}{2}$ at the first term on the right hand side of the equation is consistent choice with  $\alpha=\frac{1}{2}$ as we noticed. $\eta(x,t)$ is the Gaussian white noise field, which enjoys the same probability distribution $P(\eta)$ given in (\ref{Gaussian white noise distribution}) except the choice of $\lambda=\frac{\hbar}{2}$. Then the correlation functions between $\vec{\eta}$s are given by
\begin{equation}
\label{stochastic quantization of correlation function with white noise}
\langle \eta(x,t) \rangle = 0, \qquad \langle \eta(x,t)\eta(x',t') \rangle = \hbar \delta^{(d)}(x-x')\delta(t-t').
\end{equation}
Multi-point functions are obtained as 
\begin{equation}
\label{correlation for white gaussian noise}
\langle \eta(x_1, t_1)\cdots\eta(x_{2k},t_{2k}) \rangle =
\sum_{\text{all possible pairs of a and b}} \prod_{pairs} \langle \eta(x_a, t_a) \eta(x_b, t_b) \rangle ,
\end{equation}
while odd point correlation functions will vanish.

Correlation functions of scalar field in stochastic quantization are defined as
\begin{equation}
\label{stochastic quantization}
\langle \phi_\eta(x_1,t_1)\cdots \phi_\eta(x_k,t_k)\rangle_\eta
= \frac{\int \mathcal{D}\eta e^{-\frac{1}{2\hbar}\int d^d x dt \eta^2(x,t)} \phi_\eta(x_1,t_1)\cdots \phi_\eta(x_k,t_k)}
{\int \mathcal{D}\eta e^{-\frac{1}{2\hbar}\int d^dx dt \eta^2(x,t)}},
\end{equation}
where $\phi_\eta$ denotes the solution of the Langevin equation\eqref{stochastic field with langevin eq}. A central assertion of stochastic quantization is that in the long (stochastic) time limit, namely in its equilibrium, the stochastic correlation functions approaches Euclidean quantum correlation functions.
\begin{equation}
\lim_{t\to\infty} \langle \phi_\eta(x_1,t)\cdots \phi_\eta(x_k,t)\rangle_\eta =\langle \phi(x_1)\cdots \phi(x_k)\rangle,
\end{equation}

Solving Langevin equation will provide correlation functions of the scalar fields, but we will not discuss the detailed process in this note. This is because our main interest in this paper is Fokker-Planck approach and its Hamiltonian dynamics, even though the form of Langevin equation is very much crucial in the following discussion.


\paragraph{Probability Distribution function for scalar theory \& Fokker-Planck Lagrangian density}

Now, we construct stochastic partition function by introducing probability distribution $P[\phi,t]$ with which the $n$-point correlation function of the scalar field is defined as
\begin{equation}
\label{stochastic path integral}
\langle \phi(x_1,t)\cdots \phi(x_k,t)\rangle =\int \mathcal{D}\phi P[\phi,t] \phi(x_1)\cdots \phi(x_k),
\end{equation}
We note that the field $\phi$ does not necessarily satisfy Langevin equation and the information of the Langevin equation is already encoded in $P[\phi,t]$ as one will see. We also point out that in the long stochastic time ``$t$'' limit, the system should approach a stationary probability distribution as $\lim_{t\to\infty}P[\phi(t),t]=P_{\infty}[\phi]$, which gives the probability distribution for partition function of Euclidean theory, $S_{cl}$. 

To derive the probability distribution $P[\phi,t]$, we start with stochastic partition function,
\begin{equation}
Z_{SQ} \equiv \int \mathcal{D}\eta(x,t) \exp \left[ -\frac{1}{2\hbar}\int_{t_0}^{t} d^d x\,dt \eta^2(x,t) \right].
\end{equation}
We take variable change from the noise $\eta(x,t)$ to the field $\phi(x,t)$ using the Langevin equation.
Performing this change of variables, the partition function becomes
\begin{equation}
\label{The partition function of stochastic promote form}
Z_{SQ} =\int \mathcal{D}\phi(x,t) \det \left(\frac{\delta\eta}{\delta\phi}\right)
P[\phi,t_0] \exp \left[ -\frac{1}{2\hbar}\int_{t_0}^{t} d t' \int d^d x \left( \dot\phi(x, t')
+\frac{1}{2}\frac{\delta S_{cl}[\phi;t']}{\delta\phi(x,t')} \right)^2 \right],
\end{equation}
where the dot on top of the field at $\dot\phi(x,t')$ denotes differentiation with respect to $t'$. The initial condition for the probability distribution $P[\phi,t_0]$ can be imposed as
\begin{align}
& P[\phi,t_0] = \prod_x \delta \big(\phi(x,t_0)-\phi_{0}(x)\big)
\end{align}
where the field configuration is fixed by $\phi_{0}(x)$ for all spatial points $x$ at initial time $t_0$.
The Jacobian arising from the change of variables can be evaluated explicitly using the Langevin equation and is given by
\begin{equation}
\det\left(\frac{\delta\eta}{\delta\phi}\right) = \exp \left[ \frac{1}{4}\int_{t_0}^{t} d t'
\int d^d x \frac{\delta^2 S_{cl}[\phi;t']}{\delta\phi(x, t')^2} \right],
\end{equation}
Taking all contributions into account, the partition function can be rewritten as
\begin{align}
Z_{SQ} = & \int \mathcal{D}\phi(x,t_0)\, P(\phi,t_0)\,
e^{\frac{1}{2\hbar}S_{cl}[\phi(t_0);t_0]}\mathcal{D}\phi(x,t)\,
e^{-\frac{1}{2\hbar}S_{cl}[\phi(t);t]}\notag
\prod_{t_0<t'<t} \mathcal{D}\phi(x,t') \exp \left(-\frac{S_{FP}}{\hbar}\right)
\\
 \equiv & \int [D\phi] P(\phi(x,t),t)
\end{align}
where $P(\phi(x,t),t)$ is probability distribution and $S_{FP}$ is Fokker-Planck action, which is given by
\begin{eqnarray}
\label{Forkker-Plank_eq}
\nonumber
S_{\mathrm{FP}}(\phi(x,t),t) &=& \int d^dx\int dt\left\{\frac{1}{2}\left(\frac{\partial\phi(x,t)}{\partial t}\right)^2
+\frac{1}{8}\left(\frac{\delta S_{cl}[\phi(x,t);t]}{\delta\phi(x,t)}\right)^2
-\frac{1}{4}\hbar\frac{\delta^2 S_{cl}[\phi(x,t);t]}{\delta\phi^2(x,t)}\right\} \\ 
&-&\frac{1}{2}\int dt \frac{\partial S_{cl}[\phi(x,t);t]}{\partial t}.
\end{eqnarray}
The last term in the above expression does not appear if the classical action, $S_{cl}$ has no explicit stochastic time dependence. In many of cases, the time dependence appears only through the field, $\phi(x,t)$ and this term normally disappears.
The detailed computations are given in Appendix \ref{Appendix C : Derivation of the Jacobian and Fokker-Planck equations}.


\paragraph{Induced Fokker-Planck, Probability Conservation equation and Fokker-Planck  Hamiltonian dynamics} The equation satisfied by $P(\phi,t)$ can also be derived directly from the Langevin equation. 
The Fokker-Planck equation can be derived directly from the Langevin dynamics.
We start from the Langevin equation
\begin{equation}
\partial_t \phi(x,t) = - \frac{1}{2}\frac{\delta S_{\mathrm{cl}}}{\delta \phi(x,t)} + \eta(x,t),
\end{equation}
where the noise satisfies
\begin{equation}
\langle \eta(x,t)\eta(x',t') \rangle = \hbar \delta^{(d)}(x-x')\delta(t-t'),
\end{equation}
The probability distribution is defined by
\begin{equation}
P[\phi,t] = \int \mathrm{D\eta} P[\eta] \delta[\phi-\phi_\eta(t)]  = \langle \delta[\phi-\phi_\eta(t)] \rangle_{\eta},
\end{equation}
Taking the chain rule gives
\begin{equation}
\partial_t P[\phi,t] = - \int d^d x \frac{\delta}{\delta \phi(x)} \langle \dot{\phi}_\eta(x,t)\, \delta[\phi- \phi_\eta(t)] \rangle_{\eta},
\end{equation}
For detailed calculation details, please refer to Appendix C. 
The Fokker-Planck equation is given below
\begin{equation}
\label{Fokker-Planck Hamiltonian equation}
\partial_t P[\phi,t] = \frac{1}{2} \int d^d x \frac{\delta}{\delta \phi(x,t)}
\left(
\frac{\delta S_{\mathrm{cl}}}{\delta \phi(x,t)} + \hbar \frac{\delta}{\delta \phi(x,t)}
\right) P[\phi,t],
\end{equation}
which has the form of a continuity equation in configuration space.
Probability conservation can be expressed in the form of a continuity equation. Introducing the probability current
$J[\phi,t]$, we write
\begin{align}
&\frac{d}{dt}\int D\phi\,P[\phi,t] =
-\int D\phi\int d^dx \frac{\delta}{\delta\phi(x)}J[\phi,t], \\
&\partial_t P[\phi,t]+\int d^dx \frac{\delta}{\delta\phi(x,t)}\,J[\phi,t]=0,
\label{continuity_equation}
\end{align}
Comparing this with \eqref{Fokker-Planck Hamiltonian equation}, we identify the probability current as
\begin{equation}
J[\phi,t]\equiv -\frac{1}{2}\left(
\frac{\delta S_{\rm cl}}{\delta\phi(x,t)}P[\phi,t]
+\hbar \frac{\delta P[\phi,t]}{\delta\phi(x,t)}
\right),
\label{probability_current}
\end{equation}
Then \eqref{Fokker-Planck Hamiltonian equation} expressed as the continuity equation
\begin{equation}
\partial_t P[\phi,t] =-\int d^dx\frac{\delta}{\delta\phi(x,t)}J[\phi,t],
\end{equation}
In equilibrium, $\lim\limits_{t\rightarrow \infty}P[\phi,t]=P_{eq}[\phi]$.
Namely that the equilibrium current vanishes, $J_{eq}[\phi]=0$ for all $x$. 
We find the equilibrium distribution
\begin{equation}
P_{eq}[\phi]\propto \exp\left[-\frac{1}{\hbar}S_{cl}[\phi]\right].
\label{equilibrium_distribution}
\end{equation}
Recall the Fokker-Planck equation \eqref{Fokker-Planck Hamiltonian equation}
\begin{equation}
\frac{\partial P[\phi,t]}{\partial t} = \frac{1}{2}\int d^d x \frac{\delta}{\delta\phi(x,t)}
\left( \frac{\delta S_{cl}}{\delta\phi(x,t)} + \hbar \frac{\delta}{\delta\phi(x,t)} \right) P[\phi,t].
\end{equation}
It is convenient to express this equation in a Schr\"odinger-type form by introducing the wave function
\begin{equation}
\label{a Schrodinger-type wave function}
\Psi_S(\phi,t) \equiv P[\phi,t]e^{\frac{1}{2\hbar}S_{cl}[\phi;t]}.
\end{equation}
Requiring $\Psi_S$ to satisfy a Schr\"odinger-type evolution equation\footnote{It's important to note here that the original Schr\"odinger equation has an $i \hbar$ on the left side, but since we're using a stochastic process from above, it's natural to compare it in Euclidean form. More explicitly, solving for i is the correct way to match it with the left side of the Fokker-Planck equation.}, one obtains
\begin{equation}
\label{Fokkek-Planck Hamiltonian equation}
\frac{\partial\Psi_S(\phi,t)}{\partial t} = - \frac{1}{\hbar}\int d^d x \mathcal{H}_{\mathrm{FP}}(\frac{\delta}{\delta \phi},\phi) \Psi_S(\phi,t),
\end{equation}
where the Fokker-Planck Hamiltonian is
\begin{align}
\label{Schrodinger-type evolution operator}
\mathcal{H}_{\mathrm{FP}} 
& = \frac{1}{2}\left( -\hbar\frac{\delta}{\delta\phi(x)} +\frac{1}{ 2}\frac{\delta S_{cl}}{\delta\phi(x)} \right)
\left( \hbar\frac{\delta}{\delta\phi(x)} +\frac{1}{2}\frac{\delta S_{cl}}{\delta\phi(x)} \right) - \frac{1}{2}\frac{\partial S_{cl}[\phi;t]}{\partial t} \notag \\
& = -\frac{\hbar^{2}}{2} \frac{\delta^2}{\delta\phi(x)^2} +\frac{1}{8}\left(\frac{\delta S_{cl}}{\delta\phi(x)}\right)^2-\frac{\hbar}{4}\frac{\delta^2 S_{cl}}{\delta\phi(x)^2} - \frac{1}{2}\frac{\partial S_{cl}[\phi;t]}{\partial t}.
\end{align}
Substituting the definition \eqref{a Schrodinger-type wave function} into the Fokker–Planck equation \eqref{Fokker-Planck Hamiltonian equation} and rearranging terms, one obtains the Schrödinger-type evolution operator \eqref{Schrodinger-type evolution operator}, with the corresponding Fokker–Planck Hamiltonian.
Finally, the Fokker-Planck Lagrangian density $\mathcal{L}_{\mathrm{FP}}$ is related to the Hamiltonian $\mathcal{H}_{\mathrm{FP}}$ through a Legendre transformation.

\section{Gibbs Entropy of O(2N) Vector Model}
\label{subsec:conjecture-gkpw}
In this section, we compute Gibbs entropy of O(2N) vector model with fractional Laplacian defined in non-compact (d+1)-dimensional Euclidean space, $\mathbb R_+ \times \mathbb R^d$ in stochastic frame, where $\mathbb R_+$ is for stochastic time $0<t<\infty$.  The theory is given by
\begin{equation}
S_{cl}  =\sum_{a=1}^{2N} \int d^d k |k| \phi^a_k \phi^a_{-k},
\end{equation}
defined on d-dimensional momentum space. As we briefly mentioned in Introduction, the motivation to study this fractional Laplacian model is from holography and its reformulation with stochastic frameworks\cite{Oh:2012bx,Jatkar:2013uga,Oh:2013tsa,Oh:2015xva,Oh:2021bxx,Kim:2023bhp,Lee:2023ynb,Chae:2024zyn}.

%

%

The O(2N) vector field theory is free and so it is  collection of 2N free fields. In this calculation, we do not consider any further constraints on this theory. First we consider free fractional Laplacian scalar field and its Fokker-Planck dynamics. In the last step, we can easily generalize this to 2N- collection of scalar fields case. 

The Euclidean action that we start with is given by 
\begin{align}
S_{cl}  = \int d^d k |k| \phi_k \phi_{-k}.
\end{align}
\noindent
Once the classical action $S_{cl}$ is determined, the probability $P$ can be obtained from the Fokker-Planck equation and from this, the Gibbs entropy $S= -\int[\mathcal{D} \phi] P[\phi,t]\log \,  P[\phi,t]     = \langle -\log P \rangle$ can be obtained.
We now solve the Fokker-Planck equation with $S_{cl}$, which is given by
\begin{equation}
\label{Fokker-Planck Hamiltonian equation 2}
\partial_t P[\phi,t] = \frac{1}{2} \int d^d x \frac{\delta}{\delta \phi(x,t)}
\left(
\frac{\delta S_{\mathrm{cl}}}{\delta \phi(x,t)} + \hbar \frac{\delta}{\delta \phi(x,t)}
\right) P[\phi,t].
\end{equation}
Let us solve Fokker-Planck equation by assumption of the form of probability $P$ as

\begin{align}
\label{probability ansatz}
P &= \bar N \exp \Bigg( -\frac{1}{\hbar}\int d^d k  A(k,t) \phi_k \phi_{-k}  -\frac{1}{\hbar}\int d^d k  B(k,t) \phi_{-k}  -\frac{1}{\hbar}\int d^d k  C(k,t)   \Bigg) ,
\end{align}

\noindent where $\bar N$ is a normalization constant.
Assuming that $A(k,t)$ has a form of
\begin{align}
    A(k,t) &= \partial_t \log D(k,t),
\end{align}
makes further simplification.

Putting the ansatz into Fokker-Planck equation and comparing the coefficients of field $\phi_k$, we get the following three equations:

\begin{align}
\partial_t A(k,t) &= \partial_t \Bigg(  \partial_t (\log D(k,t)) \Bigg) = - 2\partial_t \log \Bigg( e^{-|k| t} D(k,t) \Bigg) \partial_t (\log D(-k,t)), \\
\partial_t B(k,t) &= -\partial_t \log \Bigg( e^{-|k| t} D^2 (-k,t) \Bigg) B(k,t), \\
\partial_t C(k,t) &= \hbar \partial_t \log \Bigg( e^{-|k| t} D(k,t) \Bigg) \delta(0) 
- \frac{1}{2}  B(k,t) B(-k,t).
\end{align}

The solutions are

\begin{align}
A(k,t)
=
\frac{|k|}
{1-\left(1-\pi\alpha_k\right)e^{-2|k|t}},
\end{align}

\begin{align}
B(k,t)
=
\frac{\sqrt{2\pi |k|}\,F(k)e^{-|k|t}}
{1-\left(1-\pi\alpha_k\right)e^{-2|k|t}},
\end{align}

\begin{align}
C(k,t)
={}&
\frac{\hbar}{2}
\log\left(
\frac{1-\left(1-\pi\alpha_k\right)e^{-2|k|t}}{|k|}
\right)\delta^{(d)}(0)
-\frac{\hbar}{2}G(k)\delta^{(d)}(0)
\nonumber\\
&+
\frac{\pi}{2}
\frac{
F(k)F(-k)e^{-2|k|t}
}{
1-\left(1-\pi\alpha_k\right)e^{-2|k|t}
}
-\frac12H(k).
\end{align}

\noindent where $\alpha_k,F(k),G(k),H(k)$ are arbitrary $k$ dependent functions. 

The computation of $A(k,t)$ is technically equivalent to the holographic renormalization-group treatment of a double-trace deformation. Because $\alpha_k$ controls the behavior of $A(k,t)$ near the initial state $(t \to 0)$, we regard $\alpha_k$ as the parameter that specifies the initial boundary condition.

In summary, the probability $P$ is given by

\begin{align}
P
={}&\bar N\exp\Bigg[
-\frac{1}{\hbar}\int d^dk\,
\frac{|k|}
{1-\left(1-\pi\alpha_k\right)e^{-2|k|t}}
\left(
\phi_k+
\sqrt{\frac{\pi}{2|k|}}\,
F(k)e^{-|k|t}
\right)
\left(
\phi_{-k}+
\sqrt{\frac{\pi}{2|k|}}\,
F(-k)e^{-|k|t}
\right)
\nonumber\\
&\qquad
-\frac12\int d^dk\,
\Bigg\{
\log\left(\frac{
1-\left(1-\pi\alpha_k\right)e^{-2|k|t}}{|k|}
\right)\delta^{(d)}(0)
-G(k)\delta^{(d)}(0)
-\frac{1}{\hbar}H(k)
\Bigg\}
\Bigg].
\end{align}
We note that one can redefine a new field $\bar\phi_k= \phi_k+
\sqrt{\frac{\pi}{2|k|}}\,
F(k)e^{-|k|t}$. This is just translation in the field space and so the partition function, $Z=\int \mathcal D\phi P(\phi,t) \rightarrow Z=\int \mathcal D\bar\phi P(\bar\phi,t)$ is invariant. We can set $G(k)=H(k)=0$ without loss of generality. For further computation, we will choose as such. 

Another important point is that the kernel, the coefficient of $\phi_k \phi_{-k}$ in the above equation should be remained positive for  the probability distribution to be normalizable.
To ensure this, the range of the stochastic time should be chosen properly. For the case that $1-\pi\alpha_k >0$, the range of the stochastic time needs to be $ \log(1-\pi\alpha_k)<|k|t <\infty$. 
On the other hand, for the case that $1-\pi\alpha_k <0$, the range of the stochastic time needs not to be restricted. Under such a condition, we may set $\bar N=1$. The probability $P$ is normalized at every stochastic time slice in the allowed range that we suggest above.


\paragraph{Gibbs entropy of fractional Laplacian O(2N) vector model}
\label{Entropy production of massive scalar in EAdS}

The expectation value of Gibbs entropy $S=-\int [\mathcal D\phi]P\log P=\langle -\log P\rangle $ is obtained as follows:

\begin{align}
\label{Simplified entropy ev}
 S={\int\mathcal{D}\phi (-\log P) P} =\frac{1}{2} \int d^d k  \delta^{(d)}(0) + \frac{1}{2} \int d^d k \ \log \left[ \frac{1}{2} \left(\frac{1 - (1 - \pi \alpha_k) e^{-2|k|t}}{|k|} \right) \right] \delta^{(d)}(0),
\end{align}
where $\delta^{(d)}(0)= \lim\limits_{L \rightarrow \infty}   \int_{-L/2}^{L/2} \frac{d^d x}{(2\pi)^d} e^{-i k \cdot x} \vert_{k=0}=\lim\limits_{L \rightarrow \infty}   \int_{-L/2}^{L/2} \frac{d^d x}{(2\pi)^d}=\frac{V_{d}}{(2 \pi)^d} $ , where $V_{d}$ is  called system spatial Volume.
For analytic tractability, we take $\alpha_k$ to be a constant $\alpha$. 
The expression in \eqref{Simplified entropy ev} is obtained by computing the entropy for a single fractional Laplacian scalar field in non compact flat background. 
Since the scalar fields are non-interacting, each scalar field gives an individual contribution to the entropy.
As a result, in the presence of 2N decoupled scalar fields, the total entropy is given by the sum of the individual contributions, such that
\begin{equation}
\label{Simplified entropy ev in N scalar field case}
S =\sum_{i=1}^{2N}  S_i(\alpha_i)  ,
\end{equation}
where $\alpha_i$ denotes the boundary condition parameter associated with the $i$-th scalar field.
For the simplest case, let us consider the case of two fractional Laplacian scalar fields(N=1) with different initial boundary conditions. Then the entropy is given by
\begin{equation}
 S  =S_1(\alpha_1) + S_2(\alpha_2)  ,
\end{equation}
as written explicitly in \eqref{Simplified entropy ev in N scalar field case}.
We obtain
\begin{align}
\label{Decompose and contribution in expectation value calculation result1}
S  &=  S_1(\alpha_1)  + S_2(\alpha_2)  \notag \\
&= \int d^d k \ \delta^{(d)}(0) 
+ \frac{1}{2} \delta^{(d)}(0) \int d^d k \Bigg\{ 
\log \left[ \frac{1}{2} \left(\frac{1 - (1 - \pi \alpha_1) e^{-2|k|t}}{|k|} \right) \right] \nonumber \\
&\qquad\qquad\qquad\qquad + \log \left[ \frac{1}{2} \left(\frac{1 - (1 - \pi \alpha_2) e^{-2|k|t}}{|k|} \right) \right] 
\Bigg\}.
\end{align}


Now, we discuss our specific choice of the initial boundary condition. Absolutely, there are diverse pieces of choice of boundary conditions. In that sense, O(2N)  vector model may have more degrees of freedom than large N mean field theory of O(N) vector model. Among them, our choice is an appropriate choice to reproduce the mean field limit of O(N) vector model free energy result. As we will show, the O(N) vector model we consider has a deformation parameter, $\ell$ and the theory is invariant under $\ell \rightarrow 1/\ell$ inversion. The counter part of the parameter $\ell$ is to be $\ell^{(H)}$ in O(2N) fractional Laplacian vector model and it has a relation with $\alpha_i$ as $\ell^{(H)}_i \equiv 1-\pi\alpha_i$. The boundary condition that we have chosen is given by
\begin{equation}
\ell^{(H)} \equiv \ell^{(H)}_1 = \bigl(\ell^{(H)}_2\bigr)^{-1},
\end{equation}
where $\ell^{(H)}$ is a real constant.
This also ensures that $S$ is invariant under $\ell^{(H)} \rightarrow \frac{1}{\ell^{(H)}}$ transformation as well as inversion invariant boundary conditions for the two scalar fields. 

Now we generalize this boundary condition for O(2N) vector model in such a way that for N-scalars, we request $\ell^{(H)}$ boundary condition whereas the other N-scalars, we does $1/\ell^{(H)}$ one.
Then, the form of Gibbs entropy with these 2N scalar fields is 
\begin{align}
\label{Decompose and contribution in expectation value calculation result2}
S 
&= N\int d^d k   \delta^{(d)}(0) 
+ \frac{N}{2} \delta^{(d)}(0) \int d^d k   
\log \left[ \frac{1}{4k^2} (1 - \ell^{(H)} e^{-2|k|t})\left(1 - \frac{1}{\ell^{(H)}} e^{-2|k|t} \right)\right] \notag \\
&= N\int d^d k   \delta^{(d)}(0) \Bigg[
1  - k|t| - \frac{1}{2} \log(4\ell^{(H)}) \notag \\
&\qquad\qquad\qquad\qquad 
+ \frac{1}{2} \log\left(1 + (\ell^{(H)})^2 - 2\ell^{(H)} \cosh(|k|t)\right)-\log|k|
+ \frac{\pi i}{2}
\Bigg].
\end{align}


In sum,  we start with 2N-collection of scalar fields with fractional Laplacian and so the theory has 2N degrees of freedom. The theory enjoys O(2N) global symmetry. We solve Fokker-Planck equation to get probability $P$ and it turns out that $P$ is normalized at any stochastic time slice of $t$. This means that the term $C(k,t)$ is the inverse of the logarithm of one loop determinant of the kernel, $A(k,t)$ in the solution ansatz \eqref{probability ansatz}. Therefore, the Gibbs entropy that we obtained is the one loop determinant of the kernel $C(k,r)$. As $t\rightarrow \infty$,  the Gibbs entropy approaches 2N-collection of one loop determinant of the fractional Kernel $|k|$. Then, we have
\begin{equation}
S_{t\rightarrow \infty}  \sim - N \int d^dk \log |k| \delta^{(d)}(0).
\end{equation}

\section{On calculation of free energy in O(N) vector model}
\label{On Calculation of Free Energy in CFT}
\subsection{The O(N) Non-linear sigma model}
In this section, we consider the O(N) vector model in 3 dimensions as a dual description of the holographic model that we have discussed last section. The real vector field, $\tilde{n}_{\ell}$ is SO(N) vector field where the index $\ell$ runs from 1 to N.
The vector field $\tilde{n}_{\ell}$ is designed to couple to the background gauge field  $A_{\mu}=(A_{\tau} , A_{i})=(\mu, \vec{0})$.
Especially, we discuss the simplest case where only $A_{\tau}$ is non-zero. The other component vanishes, meaning that $A_{i}=0$.
This model is the generalization of the O(N) vector model, which is in Ref. \cite{4}, where the author only discussed the case without the gauge field.
To couple with the gauge field, we promote the ordinary derivative $\partial_{\tau}$ to the covariant derivative $D_{\tau}$ where we define $D_{\tau} \equiv \partial_{\tau} - \mu$.

Now we consider a theory in the background of finite-temperature. Finite-temperature field theory can be obtained by applying periodicity to temporal coordinate. More precisely, we compactify the temporal coordinate $\tau$ and assign its periodicity $\beta$ to the vector field $\tilde{n}_{\ell}$ that has a certain boundary condition as $\tilde{n}_{\ell}(\tau + \beta) = a \tilde{n}_{\ell} ({\tau})$. For the case $a= 1$ or $a=-1$, they respectively correspond to periodic, anti-periodic boundary condition. We can also impose the twist boundary condition for $a \neq \pm 1$.

For simplicity, we consider the periodic boundary condition.
This condition demands the field $\tilde{n}_{\ell}$ have quantized frequency, $\omega_{n} =  \frac{2 \pi n}{\beta}$, where $n$ is an integer.
However, as we will see, the effective frequency becomes $\tilde{\omega}_{n} = \frac{2 \pi n}{\beta}-i\frac{\log \ell}{\beta}$, where the last term in the frequency comes from the effect of the gauge field coupled to the O(N) vector field.

The action for O(N) vector model is given by 

\begin{align}
\label{The action for O(N) vector model in large N limit}
    Z  = \int [D\Tilde{n_{l}}][D \lambda] \exp \bigg[ -\frac{1}{2g}\int d^{d} r \int_{0}^{\beta} d\tau [\tilde{n}_{l} ( - \nabla^{2} - D_{\tau}^{2} +i \lambda)\tilde{n}_{l} - i \lambda N   ]                            \bigg]  
\end{align}
\footnote{Note that $\det \hat{A} = \exp \left[ \mathrm{tr} \log \hat{A} \right]$.} Then, once we perform functional integration with respect to $\tilde{n}_{l}$, the action becomes 
\begin{align}
\label{action with lambda}
    Z = \int [D\lambda] \exp \left[
    -\frac{N}{2}  \mathrm{tr} \log \left( -D_\tau^2 - \nabla^2 + i\lambda \right) 
    + \frac{N}{2}  \mathrm{tr} \log g 
    + \frac{iN}{2g} \int d^dr \int_0^\beta d\tau  \lambda 
    \right] 
\end{align}
where $\lambda$ is an auxiliary field which restricts the magnitude of the vector field $\tilde{n}_{\ell}$ be $\sqrt N$. \\
The equation of motion of $\lambda$,
\begin{align*}
    & \frac{\delta S_{\mathrm{eff}}[\lambda]}{\delta \lambda(\vec{x}, \tau)} =0,
\end{align*}
is given by
\begin{equation}
    \label{gap eq}
    \frac{1}{g}  = \frac{1}{\beta V_d} \Tr \bigg[\frac{1}{- D_\tau^2 - \nabla^2 + m^2}\bigg],    
\end{equation}
where $i \langle \lambda \rangle = m^2$. $m^2$ is an expectation value of the field $i \lambda$.
We recall the definition of the partition function and free energy as
\begin{align}
&  Z[\lambda] := \int [D\lambda] \exp( -NS_{\mathrm{eff}}[\lambda]  ), \\
& F = -\frac{1}{\beta} \log Z,
\end{align}
where $k_B = 1$.
By comparing this to the equation \eqref{action with lambda}, we obtain the following expression of the effective action and free energy as 
\begin{align}
&S_{\text{eff}} = \frac{1}{2}   \mathrm{tr} \log \left( -D_{\tau}^{2} - \nabla^{2} + m^{2} \right)
- \frac{1}{2}   \mathrm{tr} \log g 
- \int d^{d}r \int_{0}^{\beta} d\tau   \frac{m^{2}}{2g},  
\\ & F = +\frac{1}{\beta} N S_{\text{eff}}[\lambda]
= \frac{N}{\beta} \left[
\frac{1}{2}   \mathrm{tr} \log \left( -D_{\tau}^{2} - \nabla^{2} + m^{2} \right)
- \int d^{d}r \int_{0}^{\beta} d\tau   \frac{m^{2}}{2g}
\right]. \label{free energy before evaluation}
\end{align}
The contribution $- \frac{1}{2}   \mathrm{tr} \log g $ corresponds to an additive constant (typically divergent) independent of the dynamical variables. Since only differences in the free energy are physically relevant, this constant can be subtracted.\\
We explicitly express the trace-log term in the above integration as
\begin{align*}
    \frac{1}{2}  \mathrm{tr} \log \left( -D_{\tau}^{2} - \nabla^{2} + m^{2} \right) =  \int d^{d}r \int_{0}^{\beta} d\tau  \frac{1}{2 \beta} \sum_{\tilde{\omega}_{n}} \int \frac{d^{d}\vec{k}}{(2\pi)^{d}} \log \left( \vec{k}^{2} + \tilde{\omega}_{n}^{2} + m^{2} \right),
\end{align*}
Inserting this result into \eqref{free energy before evaluation}, free energy becomes
\begin{align}
F = N V_d 
\bigg[ 
\frac{1}{2\beta} \sum_{\tilde{\omega}_n} \int \frac{d^{d}\vec{k}}{(2\pi)^d} 
\log \big( \vec{k}^{2} + \tilde{\omega}_n^2 + m^2 \big) 
- \frac{m^2}{2g}
\bigg],
\end{align}
where $\int d^{d}r \int_{0}^{\beta} d\tau$=$\beta V_d, \  V_{d}$ called system spatial Volume. 

Since the entropy is proportional to the volume and the total number of degrees of freedom diverges in the infinite-volume limit, it is useful to define free energy density rather than the free energy itself. This convention ensures that all thermodynamic quantities, such as the entropy density and energy density, remain finite. 

Then, the free energy density is given by
\begin{equation}
\label{the free energy density in momentum space}
    \frac{F}{V_d} \equiv \mathcal{F} = \frac{N}{2\beta} \sum_{\tilde{\omega}_n} \int \frac{d^{d}\vec{k}}{(2\pi)^d} \log \big( \vec{k}^{2} + \tilde{\omega}_n^2 + m^2 \big) 
- \frac{N m^2}{2g} 
\end{equation}
In the following sections, we focus on evaluating the integral and Matsubara frequency summation appearing in the free energy density expression given above.

\subsection{Decomposing the Partition Function and Reconstructing the Free Energy}
In this section, we clarify how to decompose the partition function and  the free energy step by step. Determinant is given by
\begin{equation*}
A = -D_\tau^2 +\bigg[\sqrt{- \nabla^2 + m^2}\bigg]^2
\end{equation*}
We decompose it into two operators as
\begin{equation*}
A = \bigl(-D_\tau + E \bigr) \bigl(D_\tau + E \bigr)=A_{1} A_{2},
\end{equation*}
where
\begin{equation*}
A_{1} = -D_\tau + E, \   A_{2}=D_\tau + E, \  E = \sqrt{-\nabla^2 + m^2}
\end{equation*}
This reproduces the original operator, so the determinant is
\begin{equation*}
\det A = \det(- D_\tau + E) \det( D_\tau + E) = \det A_1 \det A_2
\end{equation*}
We also decompose the partition function into $A_1$ part and $A_2$ part as
\begin{align*}
Z & = Z_1 \cdot Z_2 \\
 & = \int [D\Tilde{n_{l}}][D \lambda] \exp \bigg[ -\frac{1}{2g}\int d^{d} r \int_{0}^{\beta} d\tau [\tilde{n}_{l} ( A_1 \, A_2)\tilde{n}_{l}  - m^2 N   ]                            \bigg], \\
 &  = \int [D\lambda] \exp \left[
    -\frac{N}{2}  \mathrm{tr} \log \left( A_1\right) -\frac{N}{2}  \mathrm{tr} \log \left( A_2 \right)
    + \frac{N}{2}  \mathrm{tr} \log g 
    + \frac{ N m^2}{2g} \int d^dr \int_0^\beta d\tau   
    \right],
\end{align*}
Using the decomposition,
\begin{equation}
Z \propto \left[\det(-D_\tau + E)\right]^{-N/2} \left[\det( D_\tau + E)\right]^{-N/2},
\end{equation}
we define the decomposed partition function $Z = Z_1 \cdot Z_2$ as
\begin{equation*}
Z_1 \propto \left[\det(-D_\tau + E)\right]^{-N/2},
\quad
Z_2 \propto \left[\det( D_\tau + E)\right]^{-N/2}.
\end{equation*}
In explicit integration form, each trace-log term is given by
\begin{align*}
\frac{1}{2} \mathrm{tr}\log(-D_\tau + E)
&=  \int d^{d}r \int_{0}^{\beta} d\tau  \frac{1}{2 \beta} \sum_{\tilde{\omega}_{n}} \int \frac{d^{d}\vec{k}}{(2\pi)^{d}} \log \left( -i \tilde{\omega}_{n} + \sqrt{\vec{k}^2 + m^2} \right),  \\
\frac{1}{2} \mathrm{tr}\log( D_\tau + E) 
&=  \int d^{d}r \int_{0}^{\beta} d\tau  \frac{1}{2 \beta} \sum_{\tilde{\omega}_{n}} \int \frac{d^{d}\vec{k}}{(2\pi)^{d}} \log \left( i \tilde{\omega}_{n} + \sqrt{\vec{k}^2 + m^2} \right), 
\end{align*}
By substituting the above expression for the trace-log term into Equation \eqref{free energy before evaluation}, one obtains the free energy 
\begin{align}
\label{Decomposed Free energy density result1}
F =& F_1 + F_2 \notag \\
= &N V_d 
\bigg[ 
\frac{1}{2} \int \frac{d^{d}\vec{k}}{(2\pi)^d} \big( G \big)  - \frac{m^2}{4g}  + \frac{1}{2}   \int \frac{d^{d}\vec{k}}{(2\pi)^d} \big( \bar{G}  \big)- \frac{m^2}{4g}
\bigg],
\end{align}
We define the functions G and $\bar{G}$ as
\begin{equation*}
    \ G = \frac{1}{\beta} \sum_n \log(i\tilde{\omega}_n + \epsilon_{\vec{k}}), \ \bar{G} = \frac{1}{\beta} \sum_n \log(-i\tilde{\omega}_n + \epsilon_{\vec{k}}),
\end{equation*}
where $\epsilon_{\vec{k}}=\sqrt{ k^2 + m^2}.$ \\
By calculation, the free energy density is 
\begin{align}
\label{Decomposed Free energy density result2}
\mathcal{F}= N \bigg[ 
\frac{1}{2}  \int \frac{d^{d}\vec{k}}{(2\pi)^d} \big( G + \bar{G}  \big)- \frac{m^2}{2g}
\bigg].
\end{align}
This shows that the decomposition yields the same physical result as Equation \eqref{the free energy density in momentum space}.

\subsection{Evaluation of the Free energy density}
In this section, we evaluate the free energy density given in equation \eqref{the free energy density in momentum space}.
To evaluate the free energy density we employ the standard method in condensed matter physics theory where we define $\epsilon_{\vec{k}}=\sqrt{ k^2 + m^2}$.
One way to compute the Matsubara frequency summation is to differentiate the integrand, $\log(\tilde{\omega_{n}}^2+\epsilon^2_{\vec k})$ with respect to $\epsilon_{\vec k}$. Then, the summation becomes much easier, where we use a technique that we switch the summation to a complex variable integral. After performing the summation, we go back to the original integrand by performing integration with respect to $\epsilon_{\vec k}$. In the following, we discuss  the detailed procedure line by line.
\paragraph{Bose-Einstein distribution, $\ell=1$ case : } 
To warm up, we discuss $\ell=1$ case. 
Bose-Einstein distribution corresponds to $\ell=1$ case.
In this case, the shifted frequency, $\tilde{\omega}_{n} = \frac{2 \pi n}{\beta} - i\frac{\log \ell}{\beta}$ becomes $\omega_{n} = \frac{2 \pi n}{\beta}$. 
Now we are almost ready to compute the integrand, $\sum_{\omega_{n}}  \log({\omega_{n}}^2+\epsilon^2_{\vec k})$. Once we take derivative this with respect to $\epsilon_{\vec k}$, we have
\begin{align}
\label{Matsubara sum in BE}
\frac{1}{\beta} \sum_n \frac{2\epsilon_{\vec{k}}}{\omega_{n}^2 + \epsilon_{\vec{k}}^2} 
\end{align}
This can be understood with the aid of Figures~\ref{fig:Matsubara-contour-1}–\ref{fig:Matsubara-contour-3}.
For illustrative purposes, Figures~\ref{fig:Matsubara-contour-1}–\ref{fig:Matsubara-contour-3} show only a finite subset of the poles located along the imaginary axis, although in reality the Matsubara frequencies extend to infinity. 
These figures provide a schematic guide to the contour deformations, and the discussion below gives the details of the steps illustrated there.
To perform the Matsubara frequency summation, let us consider the following quantity:
\begin{align}
\label{Calculation of Matsubara sum}
\frac{1}{\beta} \sum_n \frac{1}{\omega_{n}^2 + \epsilon_{\vec{k}}^2} 
&=  \oint_C \frac{dz}{2\pi i} \frac{1}{-z^2 + \epsilon_{\vec{k}}^2}  \frac{1}{e^{\beta z} - 1}  = \oint_{C^{\prime}} \frac{dz}{2\pi i} (-1) \left( \frac{1}{z - \epsilon_{\vec{k}}} - \frac{1}{z + \epsilon_{\vec{k}}} \right)  \frac{1}{2\epsilon_{\vec{k}}}  \frac{1}{e^{\beta z} - 1},
\end{align}
As a result of the contour integration
\begin{align}
\label{Result of Calculation of Matsubara sum}
\frac{1}{\beta} \sum_n \frac{1}{\omega_{n}^2 + \epsilon_{\vec{k}}^2}
 & = \frac{1}{2\epsilon_{\vec{k}}} \left[ \frac{e^{-\beta \epsilon_{\vec{k}}}}{1-e^{-\beta \epsilon_{\vec{k}}}} - \frac{e^{\beta \epsilon_{\vec{k}}}}{1-e^{\beta \epsilon_{\vec{k}}}} \right]
= \frac{1}{2\epsilon_{\vec{k}}} \left[ 1 + \frac{2}{e^{\beta \epsilon_{\vec{k}}} - 1} \right].
\end{align}
The 1st equality in \eqref{Calculation of Matsubara sum} is obtained by rewriting Matsubara frequency summation as a contour integral. This is achieved by introducing the Bose-Einstein distribution function in the complex integral, which has poles at $z_0=\frac{2\pi n i}{\beta}$. Therefore, if one considers a contour enclosing the imaginary line in the complex plane, then the integral reproduces the original summation.

\begin{figure}[htbp]
\centering
\begin{minipage}[c]{0.32\textwidth}
\centering
\resizebox{\linewidth}{!}{%
\begin{tikzpicture}[scale=1.0]
\draw[->] (-3.5,0) -- (3.5,0) node[right] {$\Re$};
\draw[->] (0,-3.5) -- (0,3.5) node[above] {$\Im$};

\foreach \y in {-2.0,-1.2,0,1.2,2.0} {
  \draw[red, thick] (0.3,\y) arc[start angle=0, end angle=360, radius=0.3];
  \draw[->, red, thick] (0,\y+0.3) arc[start angle=90, end angle=135, radius=0.3];
  \node at (0,\y) {\Large $\times$};
  \node at (-0.55,\y+0.2) {$C_{i}$};
}
\draw[black, thick, postaction={decorate}, decoration={markings, mark=at position 0.975 with {\arrow{>}}}] 
  (0,0) ellipse [x radius=1.1, y radius=3.3];
\node at (1.8,2) {$C=\sum{C_{i}}$};
\node at (2,0) {\Large $\times$};
\node at (-2,0) {\Large $\times$};
\node at (2.0,0.25) {$z_{0}=\epsilon_{\vec k}$};
\node at (-2.0,0.25) {$z_{0}=-\epsilon_{\vec k}$};
\node at (0.1, -2.5) {$\vdots$};
\node at (0.1, 2.8) {$\vdots$};
\end{tikzpicture}}
\caption{Poles of the Bose–Einstein distribution on the imaginary axis(=Bosonic Matsubara frequencies) and simple poles.}
\label{fig:Matsubara-contour-1}
\end{minipage}\hfill
\begin{minipage}[c]{0.32\textwidth}
\centering
\resizebox{\linewidth}{!}{%
\begin{tikzpicture}[scale=1.0]
\def\R{2.8}
\draw[->,thin] (-3.5,0) -- (3.5,0) node[right] {$\Re$};
\draw[->,thin] (0,-3.5) -- (0,3.5) node[above] {$\Im$};

\draw[thick,postaction={decorate},decoration={markings,mark=at position 0.55 with {\arrow{<}}}] (0.5+\R,0.5) arc[start angle=0,end angle=90,radius=\R];
\draw[thick,postaction={decorate},decoration={markings,mark=at position 0.55 with {\arrow{<}}}] (-0.5,\R+0.5) arc[start angle=90,end angle=180,radius=\R];
\draw[thick,postaction={decorate},decoration={markings,mark=at position 0.55 with {\arrow{<}}}] (-\R-0.5,-0.5) arc[start angle=180,end angle=270,radius=\R];
\draw[thick,postaction={decorate},decoration={markings,mark=at position 0.55 with {\arrow{<}}}] (0.5,-\R-0.5) arc[start angle=270,end angle=360,radius=\R];

\draw[thick,
      postaction={decorate},
      decoration={markings, mark=at position 0.5 with {\arrow{>}}}
     ] (-\R-0.5,-0.5) -- (-0.5,-0.5);
     \draw[thick,
      postaction={decorate},
      decoration={markings, mark=at position 0.5 with {\arrow{>}}}
     ] (-0.5,-0.5) -- (-0.5,-0.5-\R);
     \draw[thick,
      postaction={decorate},
      decoration={markings, mark=at position 0.5 with {\arrow{<}}}
     ] (0.5,0.5) -- (\R+0.5,0.5);
     \draw[thick,
      postaction={decorate},
      decoration={markings, mark=at position 0.5 with {\arrow{>}}}
     ] (0.5,0.5) -- (0.5, \R+0.5);
     \draw[thick,
      postaction={decorate},
      decoration={markings, mark=at position 0.5 with {\arrow{<}}}
     ] (-0.5,0.5) -- (-0.5,\R+0.5);
     \draw[thick,
      postaction={decorate},
      decoration={markings, mark=at position 0.5 with {\arrow{>}}}
     ] (-0.5,0.5) -- (-\R-0.5,0.5);
     \draw[thick,
      postaction={decorate},
      decoration={markings, mark=at position 0.5 with {\arrow{>}}}
     ] (0.5,-0.5) -- (\R+0.5,-0.5);
     \draw[thick,
      postaction={decorate},
      decoration={markings, mark=at position 0.5 with {\arrow{<}}}
     ] (0.5,-0.5) -- (0.5,-\R-0.5);

\foreach \y in {-2.0,-1.2,0,1.2,2.0}{
  \node at (0,\y) {\Large $\times$};
  \node at (-0.25,\y+0.4) {$\omega_{n}$};
}

\node at (2,0) {\Large $\times$};
\node at (-2,0) {\Large $\times$};
\node at (2.2,2.2) {$\tilde{C}_1$};
\node at (-2.2,2.2) {$\tilde{C}_2$};
\node at (2.2,-2.2) {$\tilde{C}_3$};
\node at (-2.2,-2.2) {$\tilde{C}_4$};
\node at (2.0,0.25) {$z_{0}=\epsilon_{\vec k}$};
\node at (-2.0,0.25) {$z_{0}=-\epsilon_{\vec k}$};
\node at (0.1, -2.5) {$\vdots$};
\node at (0.1, 2.8) {$\vdots$};
\end{tikzpicture}
}
\caption{Decomposition of the closed contour path into the four circular sectors $\tilde{C}_{i}$ for which $(\oint_{\tilde{C}_{1}+\tilde{C}_{2}+\tilde{C}_3+\tilde{C}_4}=0)$. }
\label{fig:Matsubara-contour-2}
\end{minipage}\hfill
\begin{minipage}[c]{0.32\textwidth}
\centering
\resizebox{\linewidth}{!}{%
\begin{tikzpicture}[scale=1.0]
\draw[->] (-3.5,0) -- (3.5,0) node[right] {$\Re$};
\draw[->] (0,-3.5) -- (0,3.5) node[above] {$\Im$};
\draw[thick, postaction={decorate}, decoration={markings, mark=at position 0.975 with {\arrow{>}}}] (2.8,0) arc[start angle=0, end angle=359.9, radius=2.8];
\foreach \y in {-2.0,-1.2,0,1.2,2.0} {
  \draw[red, thick] (0.3,\y) arc[start angle=0, end angle=360, radius=0.3];
  \draw[->, red, thick] (0,\y+0.3) arc[start angle=90, end angle=135, radius=0.3];
  \node at (0,\y) {\Large $\times$};
  \node at (-0.55,\y+0.2) {$C_{i}$};
}
\draw[red, thick] (2,0) circle (0.3);
\draw[red, thick] (-2,0) circle (0.3);
\draw[->, red, thick] (2.3,0) arc[start angle=0, end angle=45, radius=0.3];
\draw[->, red, thick] (-1.7,0) arc[start angle=0, end angle=45, radius=0.3];
\node at (2,0) {\Large $\times$};
\node at (-2,0) {\Large $\times$};
\node at (2,2.4) {$ - C_{\infty}$};
\node at (2,0.6) {$C^{\prime}_{(+)}$};
\node at (-2,0.6) {$C^{\prime}_{(-)}$};
\node at (0.1, -2.5) {$\vdots$};
\node at (0.1, 2.7) {$\vdots$};
\end{tikzpicture}}
\caption{Deformation of the contour into $C'_{(+)}$ and $C'_{(-)}$ surrounding the poles at $z_0=\pm \epsilon_k$, with the large circle $C_\infty $ at infinity.}
\label{fig:Matsubara-contour-3}
\end{minipage}

\end{figure}

Now, we explain how to get the 2nd equality in the equation \eqref{Calculation of Matsubara sum}.
Let us consider a collection of contours in the complex plane.
The contours are given in Figure 1.
The integrand in \eqref{Calculation of Matsubara sum} has poles at $z_0= \pm \epsilon_{\vec{k}}$ as well as on the imaginary line, $z=\frac{2\pi n i}{\beta}$.
The contour $C=\sum{C_{i}}$ is that enclosing the poles at $z=\frac{2\pi n i}{\beta}$.
Now we want to switch the contour integration to that with another contour $C^{\prime} ={C_{+}}^{\prime}+{C_{-}}^{\prime}$, where ${C_{+}}^{\prime}$ is enclosing $z_0=\epsilon_{\vec{k}}$, ${C_{-}}^{\prime}$ is enclosing $z_0=- \epsilon_{\vec{k}}$.
To switch the locations of the contours let us consider the following collection of contours given in Figure 2.
Each circular sector has no poles inside.
Therefore, $\oint_{\tilde{C}_{1}+\tilde{C}_{2}+\tilde{C}_{3}+\tilde{C}_{4}}=0$.
However, we can deform this contour into the form given in  Figure 3.
Since, the integrand given in \eqref{Calculation of Matsubara sum} approaches zero, being proportional to $~\frac{1}{z^2}$ as $ \vert z \vert \rightarrow \infty $, thus $\int_{C_{\infty}}=0$.
Thanks to this, the integration with Matsubara frequency summation 
\footnote{Note that the sign in Equation \eqref{Calculation of Matsubara sum}, follows from taking the contour counterclockwise, which is the standard convention.}
$\oint_{C=\sum C_{i}}=-\oint_{C^{\prime} ={C_{+}}^{\prime}+{C_{-}}^{\prime}}$.
Once one performs the integration $\int_{C^{\prime}}$ one gets \eqref{Result of Calculation of Matsubara sum}.\\
As a result the equation \eqref{Result of Calculation of Matsubara sum} leads to the following expression 
\begin{align}
\label{Matsubara sum in BE}
      \frac{1}{\beta} \sum_n \frac{2\epsilon_{\vec{k}}}{\omega_{n}^2 + \epsilon_{\vec{k}}^2} & = \left( 1 + \frac{2}{e^{\beta \epsilon_{\vec{k}}} - 1} \right) 
      \end{align}
The final step is to integrate the relation \eqref{Matsubara sum in BE} with respect to $\epsilon_{\vec{k}}$ to get the original integrand,  $ \sum_{\omega_n} \log(\omega_{n}^2+\epsilon^2_{\vec k})$.
Then, finally get
\begin{align}
      \frac{1}{\beta} \sum_n \log(\omega_{n}^2 + \epsilon_{\vec{k}}^2) & = \frac{1}{\beta}  \log(2 -2 \cosh(\beta \epsilon_{\vec{k}} )) + \frac{\log(-1)}{\beta} + C_1(\beta)  ,\notag   \\  \label{Matsubara sum in BE with log form} 
      & = \frac{1}{\beta}  \log(2 -2 \cosh(\beta \epsilon_{\vec{k}} )) + \bar{C}(\beta).
\end{align}
where $C_1(\beta, \ell=1)=C_1(\beta), \ C_1(\beta) + \frac{\log(-1)}{\beta} = \bar{C}(\beta)$. The constants $C_{1}(\beta)$ and $\bar{C}(\beta)$ do not depend on $\epsilon_{\vec{k}}$. 
\paragraph{$\ell$-deformation case :}
When $\ell \neq 1$, the procedure is entirely analogous to $\ell=1$ warm-up case. 
Before discussing $\ell$-deformation case,  let us first derive $\ell$-deformed Bose-Einstein distribution starting from the (grand) canonical partition function. The $\ell$-deformed partition function is defined as
\begin{equation}
\label{deformed BE distribution}
Z_\ell = \prod_k \frac{1}{1 - \ell e^{-\beta \epsilon_{\vec{k}}}},
\end{equation}
Taking the logarithm of both sides.
\begin{equation}
\label{logarithmic partition ftn of ell deformed BE distribution}
\log Z_\ell = -\sum_k \log(1 - \ell e^{-\beta \epsilon_{\vec{k}}}),
\end{equation}
The expectation value of the energy is defined in terms of the partition function as
\begin{equation}
\langle \epsilon \rangle_\ell = -\frac{\partial}{\partial \beta} \log Z_\ell,
\end{equation}
Substituting the expression for $\log Z_\ell$.
\begin{equation}
\langle \epsilon \rangle_\ell = -\frac{\partial}{\partial \beta} \left[ -\sum_k \log(1 - \ell e^{-\beta \epsilon_{\vec{k}}}) \right]
= \sum_k \frac{\partial}{\partial \beta} \log(1 - \ell e^{-\beta \epsilon_{\vec{k}}}),
\end{equation}
We compute the derivative
\begin{equation}
\frac{\partial}{\partial \beta} \log(1 - \ell e^{-\beta \epsilon_{\vec{k}}})
= \frac{1}{1 - \ell e^{-\beta \epsilon_{\vec{k}}}}  (\ell \epsilon_{\vec{k}} e^{-\beta \epsilon_{\vec{k}}}),
\end{equation}
Therefore
\begin{equation}
\langle \epsilon \rangle_\ell = \sum_k \frac{\ell \epsilon_{\vec{k}} e^{-\beta \epsilon_{\vec{k}}}}{1 - \ell e^{-\beta \epsilon_{\vec{k}}}}
= \sum_k \epsilon_{\vec{k}}  \frac{\ell e^{-\beta \epsilon_{\vec{k}}}}{1 - \ell e^{-\beta \epsilon_{\vec{k}}}}.
\end{equation}
We define the $\ell$-deformed Bose-Einstein distribution function as
\begin{equation}
n_\ell(\epsilon_{\vec{k}}) = \frac{1}{\ell^{-1} e^{\beta \epsilon_{\vec{k}}} - 1} = \frac{\ell e^{-\beta \epsilon_{\vec{k}}}}{1 - \ell e^{-\beta \epsilon_{\vec{k}}}},
\end{equation}
Then, the energy expectation value
\begin{equation}
\langle \epsilon \rangle_\ell = \sum_k \epsilon_{\vec{k}} \  n_\ell(\epsilon_{\vec{k}}).
\end{equation}
where the deformed distribution is
\begin{equation}
\label{ell deformed bosonic like distribution}
n_\ell(\epsilon_{\vec{k}}) = \frac{1}{\ell^{-1} e^{\beta \epsilon_{\vec{k}}} - 1}.
\end{equation}
In a similar manner, the Matsubara frequency sum can be expressed as a contour integral and the evaluation
proceeds by applying the residue theorem to the simple poles of the integrand. 
In this way, the explicit expression shown in equation \eqref{Calculation of Matsubara sum by deformed} is obtained.
\begin{align}
\label{Calculation of Matsubara sum by deformed}
\frac{1}{\beta} \sum_n \frac{1}{\tilde{\omega}_n^2 + \epsilon_{\vec{k}}^2}
&= \oint_C \frac{dz}{2\pi i} \frac{1}{-z^2 + \epsilon_{\vec{k}}^2}  \frac{1}{\ell^{-1} e^{\beta z} - 1} = \frac{1}{2\epsilon_{\vec{k}}} \left[ \frac{\ell e^{-\beta \epsilon_{\vec{k}}}}{1 - \ell e^{-\beta \epsilon_{\vec{k}}}} - \frac{\ell e^{\beta \epsilon_{\vec{k}}}}{1 - \ell e^{\beta \epsilon_{\vec{k}}}} \right],
\end{align}
The equation \eqref{Matsubara sum in BE with log form by deformed} is obtained by integrating equation \eqref{Calculation of Matsubara sum by deformed} with respect to $ \epsilon_{\vec{k}} $.\begin{align}
\label{Matsubara sum in BE with log form by deformed}
 \frac{1}{\beta} \sum_n  \log(\tilde{\omega}_n^2 + \epsilon_{\vec{k}}^2) & = \frac{1}{\beta}   \left[ \log(1 - \ell e^{-\beta \epsilon_{\vec{k}}}) + \log(1 - \ell e^{\beta \epsilon_{\vec{k}}}) \right]+C_2(\beta, \ell), \\ 
 \label{ell deformed matsubara frequency summation result}
   & = \frac{1}{\beta} \left[ \log\left( \ell^2 + 1 - 2\ell \cosh(\beta \epsilon_{\vec{k}}) \right) \right] + C_{2}(\beta, \ell).
\end{align}
This is our main result in this subsection.
We note that when $\ell =1$ equation \eqref{ell deformed matsubara frequency summation result} becomes the same with equation 
\eqref{Matsubara sum in BE with log form} and so
\begin{equation}
    C_2(\beta, \ell=1) = \bar{C}(\beta)
\end{equation}
The constants $C_2(\beta, \ell)$ and $\bar{C}(\beta, \ell)$ do not depend on $\epsilon_{\vec{k}}$.
Interpretation of $\ell$-deformed Bose-Einstein distribution function may not be revealed at this section.
A detailed explanation why it takes this particular form will be provided in Section \ref{Matching result of two cals.}. \\ \\
Thus, $\ell$-deformation case, we get the free energy density is
\begin{equation}
    \mathcal{F}  = \frac{N}{2 \beta} \int \frac{d^{2}\vec{k}}{(2\pi)^2} \bigg[  \log(1 + \ell^2 - 2\ell \cosh(\beta \epsilon_{\vec{k}}))    \bigg] + \frac{N}{2} \int \frac{d^{2}\vec{k}}{(2\pi)^{2}}\bar{C}(\beta, \ell) - \frac{N m^2}{2g}.
\end{equation}

\subsection{Evaluation of the Logarithmic Sum}
In this section, we reorganize the CFT expression \eqref{Decomposed Free energy density result1} and \eqref{Decomposed Free energy density result2} in order to explicitly separate out the logarithmic structure, which will play a central role in the subsequent finite-temperature analysis.
Our theory is defined on a thermal background by compactifying the Euclidean-time direction with period $\beta$, which introduces the inverse temperature $\beta$ and naturally leads to the appearance of the Matsubara frequency summation.
The presence of two independent contributions originates from the decomposition of the theory into two real scalar sectors, which can be regarded as independent degrees of freedom in the O(N) vector model.
Correspondingly, the CFT expression \eqref{Matsubara sum in BE with log form by deformed} decomposes into two sectors, $\frac{1}{\beta} \sum_n  \log(\tilde{\omega}_n^2 + \epsilon_{\vec{k}}^2)$ = $\frac{1}{\beta} \sum_n \log(i\tilde{\omega}_n + \epsilon_{\vec{k}})$+$\frac{1}{\beta} \sum_n \log(-i\tilde{\omega}_n + \epsilon_{\vec{k}})=G + \bar{G}$. 
This choice reflects a fact that $\ell \rightarrow \frac{1}{\ell}$ under a transformation of $G \rightarrow \bar{G}$ and $\bar{G} \rightarrow G $.
We then evaluate each contribution separately using complex integral technique.
Finally, we recombine the two sectors and recover the original logarithmic structure of the CFT free energy density, while the intermediate steps are given in Equations. \eqref{Decomposed Free energy density result1} and \eqref{Decomposed Free energy density result2}.
\\

\begin{figure}[htbp]
\centering
\begin{minipage}[c]{0.32\textwidth}
\centering
\resizebox{\linewidth}{!}{%
\begin{tikzpicture}[scale=1.0]
\draw[->] (-3.5,0) -- (3.5,0) node[right] {$\Re$};
\draw[->] (0,-3.5) -- (0,3.5) node[above] {$\Im$};

\foreach \y in {-2.0,-1.2,0,1.2,2.0} {
  \draw[red, thick] (0.3,\y) arc[start angle=0, end angle=360, radius=0.3];
  \draw[->, red, thick] (0,\y+0.3) arc[start angle=90, end angle=135, radius=0.3];
  \node at (0,\y) {\Large $\times$};
  \node at (-0.55,\y+0.2) {$C_{i}$};
}
\draw[black, thick, postaction={decorate}, decoration={markings, mark=at position 0.975 with {\arrow{>}}}] 
  (0,0) ellipse [x radius=1.1, y radius=3.3];
\node at (1.8,2) {$C=\sum{C_{i}}$};
\node at (2,0) {\Large $\times$};
\node at (2.0,0.25) {$z_{0}=\epsilon_{\vec k}$};
\node at (0.1, -2.5) {$\vdots$};
\node at (0.1, 2.8) {$\vdots$};
\end{tikzpicture}}
\caption{Poles of the Bose–Einstein distribution on the imaginary axis(=Bosonic Matsubara frequencies) and simple poles.}
\label{fig:Matsubara-contour-copy 1}
\end{minipage}\hfill
\begin{minipage}[c]{0.32\textwidth}
\centering
\resizebox{\linewidth}{!}{%
\begin{tikzpicture}[scale=1.0]
\def\R{2.8}
\draw[->,thin] (-3.5,0) -- (3.5,0) node[right] {$\Re$};
\draw[->,thin] (0,-3.5) -- (0,3.5) node[above] {$\Im$};

\draw[thick,postaction={decorate},decoration={markings,mark=at position 0.55 with {\arrow{<}}}] (0.5+\R,0.5) arc[start angle=0,end angle=90,radius=\R];
\draw[thick,postaction={decorate},decoration={markings,mark=at position 0.55 with {\arrow{<}}}] (-0.5,\R+0.5) arc[start angle=90,end angle=180,radius=\R];
\draw[thick,postaction={decorate},decoration={markings,mark=at position 0.55 with {\arrow{<}}}] (-\R-0.5,-0.5) arc[start angle=180,end angle=270,radius=\R];
\draw[thick,postaction={decorate},decoration={markings,mark=at position 0.55 with {\arrow{<}}}] (0.5,-\R-0.5) arc[start angle=270,end angle=360,radius=\R];

\draw[thick,
      postaction={decorate},
      decoration={markings, mark=at position 0.5 with {\arrow{>}}}
     ] (-\R-0.5,-0.5) -- (-0.5,-0.5);
     \draw[thick,
      postaction={decorate},
      decoration={markings, mark=at position 0.5 with {\arrow{>}}}
     ] (-0.5,-0.5) -- (-0.5,-0.5-\R);
     \draw[thick,
      postaction={decorate},
      decoration={markings, mark=at position 0.5 with {\arrow{<}}}
     ] (0.5,0.5) -- (\R+0.5,0.5);
     \draw[thick,
      postaction={decorate},
      decoration={markings, mark=at position 0.5 with {\arrow{>}}}
     ] (0.5,0.5) -- (0.5, \R+0.5);
     \draw[thick,
      postaction={decorate},
      decoration={markings, mark=at position 0.5 with {\arrow{<}}}
     ] (-0.5,0.5) -- (-0.5,\R+0.5);
     \draw[thick,
      postaction={decorate},
      decoration={markings, mark=at position 0.5 with {\arrow{>}}}
     ] (-0.5,0.5) -- (-\R-0.5,0.5);
     \draw[thick,
      postaction={decorate},
      decoration={markings, mark=at position 0.5 with {\arrow{>}}}
     ] (0.5,-0.5) -- (\R+0.5,-0.5);
     \draw[thick,
      postaction={decorate},
      decoration={markings, mark=at position 0.5 with {\arrow{<}}}
     ] (0.5,-0.5) -- (0.5,-\R-0.5);

\foreach \y in {-2.0,-1.2,0,1.2,2.0}{
  \node at (0,\y) {\Large $\times$};
  \node at (-0.25,\y+0.4) {$\omega_{n}$};
}

\node at (2,0) {\Large $\times$};
\node at (2.2,2.2) {$\tilde{C}_1$};
\node at (-2.2,2.2) {$\tilde{C}_2$};
\node at (2.2,-2.2) {$\tilde{C}_3$};
\node at (-2.2,-2.2) {$\tilde{C}_4$};
\node at (2.0,0.25) {$z_{0}=\epsilon_{\vec k}$};
\node at (0.1, -2.5) {$\vdots$};
\node at (0.1, 2.8) {$\vdots$};
\end{tikzpicture}
}
\caption{Decomposition of the closed contour path into the four circular sectors $\tilde{C}_{i}$ for which $(\oint_{\tilde{C}_{1}+\tilde{C}_{2}+\tilde{C}_3+\tilde{C}_4}=0)$. }
\label{fig:Matsubara-contour-copy 2}
\end{minipage}\hfill
\begin{minipage}[c]{0.32\textwidth}
\centering
\resizebox{\linewidth}{!}{%
     \centering
    \resizebox{0.42\linewidth}{!}{%
        \begin{tikzpicture}[scale=1.0]

            \draw[->] (-3.5,0) -- (3.5,0) node[right] {$\Re$};
            \draw[->] (0,-3.5) -- (0,3.5) node[above] {$\Im$};

            \draw[thick, postaction={decorate}, decoration={
                markings,
                mark=at position 0.975 with {\arrow{>}}
            }] (2.8,0) arc[start angle=0, end angle=359.9, radius=3.0];

            \foreach \y in {-2.0,-1.2,0,1.2,2.0} {
                \draw[red, thick] (0.3,\y) arc[start angle=0, end angle=360, radius=0.3];
                \draw[->, red, thick] (0,\y+0.3) arc[start angle=90, end angle=135, radius=0.3];
                \node at (0,\y) {\Large $\times$};
                \node at (-0.55,\y+0.3) {$C_{i}$};
            }

            \draw[red, thick] (2,0) circle (0.3);
            \draw[->, red, thick] (2.3,0) arc[start angle=0, end angle=45, radius=0.3];
            \node at (2,0) {\Large $\times$};
            \node at (2,2.5) {$-C_{\infty}$};
            \node at (1.5,0.2) {$C^{\prime}$};
            \node at (0.1, -2.5) {$\vdots$};
            \node at (0.1, 2.8) {$\vdots$};

        \end{tikzpicture}
    }
    }
    \caption{Schematic of the total contour ; the large contour goes around \textbf{clockwise}, while the small red circles, the Matsubara frequencies and simple pole, go around \textbf{counterclockwise}.}
    \label{fig:total-contour} 
\end{minipage}

\end{figure}

\noindent Let us evaluate G.
\begin{equation}
\label{decomposition of logarithmic contribution1}
G = \frac{1}{\beta} \sum_n  \log(i\tilde{\omega}_n + \epsilon_{\vec{k}})    
\end{equation}
Differentiate the above equation with respect to $\epsilon_{\vec{k}}.$
\begin{align*}
\frac{1}{\beta} \sum_n \frac{\partial}{\partial\epsilon_{\vec{k}}}  \log(i\tilde{\omega}_n + \epsilon_{\vec{k}})  & =
\frac{1}{\beta} \sum_n \frac{1}{i\tilde{\omega}_n + \epsilon_{\vec{k}}},
\end{align*}
Using the Matsubara frequency summation. 
\begin{align*}
\frac{1}{\beta} \sum_n \frac{1}{i\tilde{\omega}_n + \epsilon_{\vec{k}}}
& = \oint_C \frac{dz}{2 \pi i} (-1) \frac{1}{z - \epsilon_{\vec{k}}}  \frac{1}{\ell^{-1} e^{\beta z} - 1}, 
\end{align*}
As shown in Figure \ref{fig:total-contour}, the integration contour is chosen to enclose both the simple pole at $z_{0}$ = $\epsilon_{\vec{k}} $ and the Matsubara poles on the imaginary axis. The contour integral can be expressed as follows.
\begin{align*}
 \oint_C \frac{dz}{2 \pi i} (-1) \frac{1}{z - \epsilon_{\vec{k}}}  \frac{1}{\ell^{-1} e^{\beta z} - 1}  & = - \Res[(-1)\frac{1}{z - \epsilon_{\vec{k}}}  \frac{1}{\ell^{-1} e^{\beta z} - 1}, z_{0}=\epsilon_{\vec{k}} ] + \int_{C_{\infty}} \frac{dz}{2 \pi i} (-1) \frac{1}{z -  \epsilon_{\vec{k}}}  \frac{1}{\ell^{-1} e^{\beta z} - 1},
\end{align*}
Next, the contour integral is evaluated by applying the residue theorem. The integrand has a simple pole at $z_{0}$ = $\epsilon_{\vec{k}}$
and the corresponding residue gives the first term on the right-hand side.
The last term comes from the contour integral along $C_\infty$
which represents the contribution from the large circle at infinity. Since the integrand decays as $\frac{1}{z}$ for $Re(z)>0, \ \vert z \vert \rightarrow \infty$ is zero value but $Re(z)<0, \ Re(z) \rightarrow -\infty$ is valid value $+\frac{1}{2}$.
The following result is 
\begin{equation*}
    \oint \frac{dz}{2\pi i} (-1) \frac{1}{z - \epsilon_{\vec{k}}} \frac{1}{\ell^{-1} e^{\beta z} - 1}
= \frac{1}{\ell^{-1} e^{\beta \epsilon_{\vec{k}}} - 1} + \frac{1}{2},
\end{equation*}
Integrating the above equation with respect to $\epsilon_{\vec{k}}$, we obtain
\begin{equation}
\label{decomposition logarithmic part by ell1}
    G  = \frac{1}{2} \epsilon_{\vec{k}} + \frac{1}{\beta} \log\left(1 - \ell e^{-\beta \epsilon_{\vec{k}}} \right)  + \tilde{C}_{1}(\beta, \ell).
\end{equation}
The constant $\tilde{C}_{1}(\beta, \ell)$ does not depend on $\epsilon_{\vec{k}}$.
\begin{figure}[htbp]
\centering
\begin{minipage}[c]{0.32\textwidth}
\centering
\resizebox{\linewidth}{!}{%
\begin{tikzpicture}[scale=1.0]
\draw[->] (-3.5,0) -- (3.5,0) node[right] {$\Re$};
\draw[->] (0,-3.5) -- (0,3.5) node[above] {$\Im$};

\foreach \y in {-2.0,-1.2,0,1.2,2.0} {
  \draw[red, thick] (0.3,\y) arc[start angle=0, end angle=360, radius=0.3];
  \draw[->, red, thick] (0,\y+0.3) arc[start angle=90, end angle=135, radius=0.3];
  \node at (0,\y) {\Large $\times$};
  \node at (-0.55,\y+0.2) {$C_{i}$};
}
\draw[black, thick, postaction={decorate}, decoration={markings, mark=at position 0.975 with {\arrow{>}}}] 
  (0,0) ellipse [x radius=1.1, y radius=3.3];
\node at (1.8,2) {$C=\sum{C_{i}}$};
\node at (-2,0) {\Large $\times$};
\node at (-2.1,0.25) {$z_{0}=-\epsilon_{\vec k}$};
\node at (0.1, -2.5) {$\vdots$};
\node at (0.1, 2.8) {$\vdots$};
\end{tikzpicture}}
\caption{Poles of the Bose–Einstein distribution on the imaginary axis(=Bosonic Matsubara frequencies) and simple poles.}
\label{fig:Matsubara-contour-copy 1}
\end{minipage}\hfill
\begin{minipage}[c]{0.32\textwidth}
\centering
\resizebox{\linewidth}{!}{%
\begin{tikzpicture}[scale=1.0]
\def\R{2.8}
\draw[->,thin] (-3.5,0) -- (3.5,0) node[right] {$\Re$};
\draw[->,thin] (0,-3.5) -- (0,3.5) node[above] {$\Im$};

\draw[thick,postaction={decorate},decoration={markings,mark=at position 0.55 with {\arrow{<}}}] (0.5+\R,0.5) arc[start angle=0,end angle=90,radius=\R];
\draw[thick,postaction={decorate},decoration={markings,mark=at position 0.55 with {\arrow{<}}}] (-0.5,\R+0.5) arc[start angle=90,end angle=180,radius=\R];
\draw[thick,postaction={decorate},decoration={markings,mark=at position 0.55 with {\arrow{<}}}] (-\R-0.5,-0.5) arc[start angle=180,end angle=270,radius=\R];
\draw[thick,postaction={decorate},decoration={markings,mark=at position 0.55 with {\arrow{<}}}] (0.5,-\R-0.5) arc[start angle=270,end angle=360,radius=\R];

\draw[thick,
      postaction={decorate},
      decoration={markings, mark=at position 0.5 with {\arrow{>}}}
     ] (-\R-0.5,-0.5) -- (-0.5,-0.5);
     \draw[thick,
      postaction={decorate},
      decoration={markings, mark=at position 0.5 with {\arrow{>}}}
     ] (-0.5,-0.5) -- (-0.5,-0.5-\R);
     \draw[thick,
      postaction={decorate},
      decoration={markings, mark=at position 0.5 with {\arrow{<}}}
     ] (0.5,0.5) -- (\R+0.5,0.5);
     \draw[thick,
      postaction={decorate},
      decoration={markings, mark=at position 0.5 with {\arrow{>}}}
     ] (0.5,0.5) -- (0.5, \R+0.5);
     \draw[thick,
      postaction={decorate},
      decoration={markings, mark=at position 0.5 with {\arrow{<}}}
     ] (-0.5,0.5) -- (-0.5,\R+0.5);
     \draw[thick,
      postaction={decorate},
      decoration={markings, mark=at position 0.5 with {\arrow{>}}}
     ] (-0.5,0.5) -- (-\R-0.5,0.5);
     \draw[thick,
      postaction={decorate},
      decoration={markings, mark=at position 0.5 with {\arrow{>}}}
     ] (0.5,-0.5) -- (\R+0.5,-0.5);
     \draw[thick,
      postaction={decorate},
      decoration={markings, mark=at position 0.5 with {\arrow{<}}}
     ] (0.5,-0.5) -- (0.5,-\R-0.5);

\foreach \y in {-2.0,-1.2,0,1.2,2.0}{
  \node at (0,\y) {\Large $\times$};
  \node at (-0.25,\y+0.4) {$\omega_{n}$};
}

\node at (-2,0) {\Large $\times$};
\node at (2.2,2.2) {$\tilde{C}_1$};
\node at (-2.2,2.2) {$\tilde{C}_2$};
\node at (2.2,-2.2) {$\tilde{C}_3$};
\node at (-2.2,-2.2) {$\tilde{C}_4$};
\node at (-2.1,0.25) {$z_{0}=-\epsilon_{\vec k}$};
\node at (0.1, -2.5) {$\vdots$};
\node at (0.1, 2.8) {$\vdots$};
\end{tikzpicture}
}
\caption{Decomposition of the closed contour path into the four circular sectors $\tilde{C}_{i}$ for which $(\oint_{\tilde{C}_{1}+\tilde{C}_{2}+\tilde{C}_3+\tilde{C}_4}=0)$. }
\label{fig:Matsubara-contour-copy 2}
\end{minipage}\hfill
\begin{minipage}[c]{0.32\textwidth}
\centering
\resizebox{\linewidth}{!}{%
     \centering
    \resizebox{0.42\linewidth}{!}{%
        \begin{tikzpicture}[scale=1.0]

            \draw[->] (-3.5,0) -- (3.5,0) node[right] {$\Re$};
            \draw[->] (0,-3.5) -- (0,3.5) node[above] {$\Im$};

            \draw[thick, postaction={decorate}, decoration={
                markings,
                mark=at position 0.975 with {\arrow{>}}
            }] (2.8,0) arc[start angle=0, end angle=359.9, radius=3.0];

            \foreach \y in {-2.0,-1.2,0,1.2,2.0} {
                \draw[red, thick] (0.3,\y) arc[start angle=0, end angle=360, radius=0.3];
                \draw[->, red, thick] (0,\y+0.3) arc[start angle=90, end angle=135, radius=0.3];
                \node at (0,\y) {\Large $\times$};
                \node at (-0.55,\y+0.3) {$C_{i}$};
            }

            \draw[red, thick] (-2,0) circle (0.3);
            \draw[->, red, thick] (-1.7,0) arc[start angle=0, end angle=45, radius=0.3];
            \node at (-2,0) {\Large $\times$};
            \node at (2,2.5) {$ -C_{\infty}$};
            \node at (-1.5,0.4) {$C^{\prime}$};
            \node at (0.1, -2.5) {$\vdots$};
            \node at (0.1, 2.8) {$\vdots$};

        \end{tikzpicture}
    }
    }
    \caption{Schematic of the total contour ; the large contour goes around \textbf{clockwise}, while the small red circles, the Matsubara frequencies and simple pole, go around \textbf{counterclockwise}.}
    \label{fig:total-contour1}
\end{minipage}

\end{figure}
\\
We now follow the same procedure as before. Let us evaluate $\bar{G}$.
\begin{equation}
\label{decomposition of logarithmic contribution1}
\bar{G} = \frac{1}{\beta} \sum_n  \log(-i\tilde{\omega}_n + \epsilon_{\vec{k}})    
\end{equation}
Differentiate the above equation with respect to $\epsilon_{\vec{k}}.$
\begin{align*}
\frac{1}{\beta} \sum_n \frac{\partial}{\partial\epsilon_{\vec{k}}} \log(-i\tilde{\omega}_n + \epsilon_{\vec{k}}) 
&= \frac{1}{\beta} \sum_n \frac{1}{-i\tilde{\omega}_n + \epsilon_{\vec{k}}}, 
\end{align*}
Using the Matsubara frequency summation.
\begin{align*}
\frac{1}{\beta} \sum_n \frac{1}{-i\tilde{\omega}_n + \epsilon_{\vec{k}}}
&= \oint \frac{dz}{2 \pi i} \frac{1}{z + \epsilon_{\vec{k}}} \frac{1}{ \ell^{-1} e^{\beta z} - 1}, \\
\end{align*}
Similarly, as shown in Figure~\ref{fig:total-contour1}, the integration contour is chosen to enclose both the simple pole at $z_{0}$ = $-\epsilon_{\vec{k}} $ and the Matsubara poles on the imaginary axis. The contour integral can be expressed as follows.
\begin{align*}
\oint \frac{dz}{2 \pi i} \frac{1}{z + \epsilon_{\vec{k}}} \frac{1}{ \ell^{-1} e^{\beta z} - 1}     &  = -\Res[\frac{1}{z + \epsilon_{\vec{k}}} \frac{1}{ \ell^{-1} e^{\beta z} - 1} , z_{0}=-\epsilon_{\vec{k}} ] + \int_{C_{\infty}} \frac{dz}{2 \pi i} \frac{1}{z + \epsilon_{\vec{k}}}  \frac{1}{\ell^{-1} e^{\beta z} - 1},
\end{align*}
Thus, the residue calculation leads to the following result
\begin{align*}
\oint \frac{dz}{2 \pi i}  \frac{1}{z + \epsilon_{\vec{k}}} \frac{1}{ \ell^{-1} e^{\beta z} - 1}    & = \frac{-1}{\ell^{-1} e^{-\beta \epsilon_{\vec{k}}} - 1} - \frac{1}{2}, 
\end{align*}
Integrating the above equation with respect to $\epsilon_{\vec{k}}$, we obtain
\begin{align}
\label{decomposition logarithmic part by ell2}
\bar{G}= -\frac{1}{2} \epsilon_{\vec{k}} + \frac{1}{\beta} \log\left(1 - \ell e^{\beta \epsilon_{\vec{k}}} \right) + \tilde{C}_{2}(\beta, \ell).
\end{align}
The constant $\tilde{C}_{2}(\beta, \ell)$ does not depend on $\epsilon_{\vec{k}}$. \\
Recall $G \vert_{\ell \rightarrow \frac{1}{\ell}} \Rightarrow \bar{G}$, by looking at the definitions of G and $\bar{G}.$
Taking the inversion transformation, $\ell \rightarrow \frac{1}{\ell}$, to \eqref{decomposition logarithmic part by ell1} precisely reproduces \eqref{decomposition logarithmic part by ell2}. 
This demands that should be $\bar{G}$.
\begin{align}
\label{decomposition logarithmic conjugation realtion}
 G \vert_{\ell \rightarrow \frac{1}{\ell}} = \frac{1}{2}\epsilon_{\vec{k}} + \frac{1}{\beta}\log\left(1-\frac{1}{\ell} e^{-\beta \epsilon_{\vec{k}}}\right) + \tilde{C}_{1}(\beta,\frac{1}{\ell}),
\end{align}
Therefore
\begin{align}
G \vert_{\ell \rightarrow \frac{1}{\ell}} & 
\label{decomposition logarithmic conjugation result ell1}
= -\frac{1}{2}\epsilon_{\vec{k}} + \frac{1}{\beta}\log\left(1-\ell e^{\beta \epsilon_{\vec{k}}}\right) - \frac{1}{\beta} \log(\ell)+\frac{1}{\beta} \log(-1) + \tilde{C}_{1}(\beta,\frac{1}{\ell}) = \bar{G}.
\end{align}
Hence, the relation between G and $\bar{G}$ under the transformation $\ell \rightarrow \frac{1}{\ell}$ implies that the associated constants must satisfy the following relation.
\begin{equation*}
    - \frac{1}{\beta} \log(\ell)+\frac{1}{\beta} \log(-1) + \tilde{C}_{1}(\beta,\frac{1}{\ell}) = \tilde{C}_{2}(\beta, \ell),
\end{equation*}
The constants $\tilde{C}_{1}(\beta,\ell)$ and $\tilde{C}_{2}(\beta,\ell)$, although dependent on $\beta$ and $\ell$, are independent of $ \epsilon_{\vec{k}} $.
The constant obtained on the CFT side will later play an important role in determining the difference in the free energy density per degree of freedom, which originates from the difference in boundary conditions. 
We define the following constant relation
\begin{equation}
    \tilde{C}_1(\beta, \ell) + \tilde{C}_2(\beta, \ell) = \tilde{C}(\beta, \ell),
\end{equation}
By applying the above definition, we obtain
\begin{equation}
\label{tilde constant relation}
    -\frac{1}{\beta}\log(\ell) + A(\beta, \ell)=\tilde{C}(\beta, \ell),
\end{equation}
where $A(\beta, \ell) = \frac{1}{\beta}\log(-1)
    + \tilde{C}_1\left(\beta, \frac{1}{\ell}\right)
    + \tilde{C}_1(\beta, \ell).$ \\
Also, $A(\beta, \ell) = A(\beta, \frac{1}{\ell})$. If the constant takes the form
\begin{equation}
    \tilde{C}(\beta, \ell_1) - \tilde{C}(\beta, \ell_2)
    = \frac{1}{\beta}\log\left( \frac{\ell_2}{\ell_1} \right)+ A(\beta, \ell_1) - A(\beta, \ell_2) .
\end{equation}
this relation will be further analyzed in the context of the AdS/CFT correspondence 
and discussed in detail in Section \ref{Matching result of two cals.}.

Before moving on, let us emphasize  the results obtained above.
The two expressions G and $\bar{G}$ obtained above are related by $\mathbb{Z}_2$ invariance(i.e $\log \ell \rightarrow -\log \ell$) under the $\ell \rightarrow \frac{1}{\ell}$ transformation, as shown explicitly in equation \eqref{decomposition logarithmic conjugation result ell1}. 
An important consequence of this relation is that the combination $(G+\bar{G})\vert_{\ell \rightarrow \frac{1}{\ell}}= \bar{G}+G$ is invariant under the $\ell \rightarrow \frac{1}{\ell}$ transformation. 
Since all parameters involved in $G$ and $\bar{G}$ are real quantities ($\ell \in \mathbb{R} , \beta \in \mathbb{R^{+}}, \epsilon_{\vec{k}} \in \mathbb{R^{+}}$), both functions are real valued mappings, $G, \bar{G}: \mathbb{R} \times \mathbb{R^{+}} \times \mathbb{R^{+}} \to \mathbb{R},$ and therefore satisfy $G^{*} = G, \ \bar{G}^{*} = \bar{G}.$
Especially, $\ell = \frac{1}{\ell}=1$ is a fixed point of this transformation which corresponds to Bose-Einstein case. 
This $\ell \leftrightarrow \frac{1}{\ell}$ duality will play a central role in the holographic comparison discussed in the Section \ref{Matching result of two cals.}.

\subsection{Free energy density in O(N) vector model}
Having evaluated the contributions from each part separately, we now recombine them to recover the original logarithmic structure. 
Specifically, the two terms obtained from the decomposition add up to the full Matsubara frequency summation,
\begin{equation}
    \frac{1}{\beta}\sum_{n}\log(\tilde{\omega}_n^2+\epsilon_k^2),
\end{equation}
This recombination step is essential, since it demonstrates that the part-by-part decomposition does not lose any information and precisely reproduces the original logarithmic form. \\
By comparing this with equations \eqref{decomposition logarithmic part by ell1} and \eqref{decomposition logarithmic part by ell2}, we obtain equation \eqref{Original logarithmic form of matsubar frequency sum with ell deformed}.
\begin{align}
\label{Original logarithmic form of matsubar frequency sum with ell deformed}
\frac{1}{\beta} \sum_n \log(\tilde{\omega}_n^2 + \epsilon_{\vec{k}}^2)& = \frac{1}{\beta} \log(1 - \ell e^{-\beta \epsilon_{\vec{k}}}) 
+ \frac{1}{\beta} \log(1 - \ell e^{\beta \epsilon_{\vec{k}}})  + \tilde{C}(\beta, \ell), 
\end{align}
where $\tilde{C}(\beta, \ell) = -\frac{1}{\beta}\log(\ell) + A(\beta, \ell), \ A(\beta, \ell) = A(\beta, \frac{1}{\ell})$. The constant does not depend on $\epsilon_{\vec{k}}$. \\
Recall the Free energy density \eqref{the free energy density in momentum space} given by 
\begin{equation*}
    \frac{F}{V_d} \equiv \mathcal{F} 
= \frac{N}{2\beta} \sum_{\omega_n} \int \frac{d^{d}\vec{k}}{(2\pi)^d} 
\log \big( \vec{k}^{2} + \omega_n^2 + m^2 \big)
- \frac{N m^2}{2g},
\end{equation*}
Considering the dispersion relation $\vec{k}^{2}+m^2 = \epsilon_{\vec{k}}^2$
\begin{equation}
     \mathcal{F} = \frac{N}{2} \frac{1}{\beta}\sum_{\omega_n} \int \frac{d^{d}\vec{k}}{(2\pi)^d} 
\log \big(  \omega_n^2 +  \epsilon_{\vec{k}}^2 \big) 
- \frac{N m^2}{2g},
\end{equation}
Particularly for the Bosonic case($\ell=1$), substituting the result\footnote{See Appendix A for further details.} \eqref{Matsubara sum in BE with log form} into \eqref{the free energy density in momentum space} given by 
\begin{equation}
\label{Reduced Bosonic free energy density}
     \mathcal{F}  = \frac{N}{2\beta} \int \frac{d^{d}\vec{k}}{(2\pi)^{d}}  \bigg[    \log(2 -2 \cosh(\beta \epsilon_{\vec{k}} ))   \bigg] +  \frac{N}{2} \int \frac{d^{d}\vec{k}}{(2\pi)^{d}} \bar{C}(\beta)
- \frac{N m^2}{2g},
\end{equation}
By substituting the result \eqref{Original logarithmic form of matsubar frequency sum with ell deformed} into \eqref{Decomposed Free energy density result2}, we obtain the following result
\begin{equation}
    \frac{\mathcal{F}}{N}  = \frac{1}{2 \beta} \int \frac{d^{d}\vec{k}}{(2\pi)^d} \bigg[  \log(1 + \ell^2 - 2\ell \cosh(\beta \epsilon_{\vec{k}}))    \bigg] + \frac{1}{2} \int \frac{d^{d}\vec{k}}{(2\pi)^{d}}\tilde{C}(\beta, \ell) - \frac{ m^2}{2g}.
\end{equation}
When we set $\ell = 1$ in the $\ell$-deformed free energy expression, the result precisely reduces to the bosonic free energy density. 
The integrands match exactly up to the constant terms, provided \eqref{Reduced Bosonic free energy density} that the constants $\bar{C}$ and $\tilde{C}$ are equal. 
Since both $\bar{C}$ and $\tilde{C}$ are independent of $\epsilon_{\vec{k}}$, they do not affect the momentum dependent structure of the free energy.
In this work, we normalize the free energy density by the number of degrees of freedom $N$, and define $\frac{\mathcal{F}}{N}=\frac{\sum_i F_{i}}{N}$. This quantity can be interpreted as the free energy density per degree of freedom.
Before turning to Section \ref{Matching result of two cals.}, let us look at the structure that has emerged on the CFT side. 
As we have seen, the two contributions $G$ and $\bar{G}$ are related by $\ell \rightarrow \frac{1}{\ell}$ transformation, and their sum $G+\bar{G}$ is invariant under the $\ell \rightarrow \frac{1}{\ell}$. 

\section{Matching}
\subsection{Matching Between $O(N)$ vector model free-energy and Gibbs entropy of O(2N) Vector Model} 
\label{Matching result of two cals.}
In this section, we will discuss a noticeable relation between the two different results from the previous sections. One is free-energy calculation from $O(N)$ vector model in $d+1$-dimension. The spacetime on which the theory is defined is $S^1\times \mathbb R_d$, where the $S^1$ is the thermal circle. The theory is obtained by deforming the usual $O(N)$ vector model by employing
a deformation $\mu=\beta^{-1}\log l$,
where the $l$ is the deformation parameter to the original theory. As the usual process to deal with this theory, we consider an auxiliary field, $\lambda$ to preserve the magnitude of the $O(N)$ vector field as $\tilde n^a\tilde n^a=N$. Inserting this constraint with a form of delta function and exponentiating it provide Gaussian(functional) integration with the $O(N)$ vector field in the partition function. This gives one-loop determinant of $O(N)$ vector field model, where the auxiliary field $\lambda$ is staying in. We assume that the expectation value of the field $\lambda$ has a form of $i\langle\lambda\rangle=m^2$, where $m$ is a constant. We concentrate on the thermal average of the theory, and we take perform the possible frequency summation, so called Matsubara frequency summation, on the imaginary time circle. Finally, we take logarithm of the partition function to get free energy of the theory. 

Another result is from O(2N) vector model with fractional Laplacian kernel. Fractional Laplacian theory frequently appears as a generating functional of boundary field theory of bulk scalar field model in holography. Moreover, the holographic framework is successfully reformulated with stochastic dynamics as explained in introduction. With such a motivation, we perform Fokker-Planck dynamics with this O(2N) fractional Laplacian vector model.
In our context, we consider 2N collection of scalar fields with a boundary condition in such a way that half of them show different boundary condition against the other half.
%
The boundary condition is carried by a parameter $\ell^{(H)}_a$, where $a=1$ to $2N$. We assign the boundary conditions as
$\ell^{(H)}_{a} = \ell^{(H)}$ for $a=1$ to $N$ and $\ell^{(H)}_{a} = \frac{1}{\ell^{(H)}}$ for $a=N+1$ to $2N$.

Let us pause  and explain why we consider such a boundary condition. In the $l$-deformed $O(N)$ vector model, there is an interesting symmetry that under the transform $\ell \rightarrow \frac{1}{\ell}$, the free-energy is invariant. $\ell$ is the deformation parameter and in the holographic context, the deformation in the dual field theory is related to boundary condition of the fields in the dual gravity model. Because holography can be reformulated with stochastic dynamics, the latter also reflects such properties of the former. Since the deformation parameter has the inversion symmetry, it is very natural the boundary condition perceive the symmetry in the holographic dual gravity model.

The entropy expectation value in the bulk can be computed via stochastic quantization techniques, as shown in Section \ref{Entropy production of massive scalar in EAdS}. 
The entropy expectation value can be written by
\begin{equation*}
 S  = N\int d^d k \delta^{(d)}(0) \left[ 1  - \vert k \vert t - \frac{1}{2} \log(4|k|^2 (\ell^{(H)})) + \frac{1}{2} \log(1 + (\ell^{(H)})^2 - 2 (\ell^{(H)}) \cosh(2 \vert k \vert t) ) + \frac{\pi i}{2} \right],
\end{equation*}
$\delta^{(d)}(0) = V_{d}/ (2 \pi)^d$, $V_{d}$ called system spatial Volume. We also define the entropy density as $s = S /V_{d}$.
\paragraph{Comparison of the results from the two theories}
Now, we compare the two results from the previous section:
One is $\ell$ deformed O(N) vector model calculation, 
\begin{align}
& \frac{\mathcal{F}}{N} = \frac{1}{2 \beta} \int \frac{d^{d}\vec{k}}{(2\pi)^d} \bigg[  \log(1 + \ell^2 - 2\ell \cosh(\beta \epsilon_{\vec{k}}))    \bigg] + \frac{1}{2} \int \frac{d^{d}\vec{k}}{(2\pi)^{d}}\tilde{C}(\beta, \ell) - \frac{ m^2}{2g},
\end{align}
another is O(2N) model Gibbs entropy, 
\begin{equation}
 s  = N \int \frac{d^d k}{(2\pi)^d}   \left[  1  - \vert k \vert t  - \frac{1}{2} \log(4 |k|^2(\ell^{(H)}) ) + \frac{1}{2} \log(1 + (\ell^{(H)})^2 - 2(\ell^{(H)}) \cosh(2 \vert k \vert t)) + \frac{\pi i}{2} \right],
\end{equation}

To compare these each other, we regularize them as follows. We regularize the Gibbs entropy by subtracting it with the same quantity at $t=\infty$.
Then the regularized Gibbs entropy density is given by
\begin{align}
\label{Holographic cal}
    \Delta s  &=  s(\ell^{(H)},t) -  s(\ell^{(H)}, \infty), \notag \\
    &= N \int \frac{d^d k}{(2\pi)^d}   \left[  - \vert k \vert t  - \frac{1}{2} \log( \ell^{(H)})  + \frac{1}{2} \log(1 + (\ell^{(H)})^2 - 2(\ell^{(H)}) \cosh(2 \vert k \vert t)) + \frac{\pi i}{2}\right].
\end{align}
We note that $\langle s(\ell^{(H)},\infty) \rangle=N \int \frac{d^d k}{(2\pi)^d} (1-\log 2-\log |k|)$, and it reflects the late time behavior of the stochastic dynamics of O(2N) vector model where $-N\log|k|$ indicates the dynamical degrees of freedom of the model. 
In the similar way, the free energy density of O(N) vector model needs to be regularized. then, we have
\begin{align}
\beta {\Delta\mathcal{F}}\equiv\beta \bigg[ 
{\mathcal{F}(\ell,\beta)-\mathcal{F}(\ell,\infty)}
\bigg]
&= N\int \frac{d^d k}{(2\pi)^d}
\left[
\frac12 \log \big(1+\ell^2-2\ell\cosh(\beta\epsilon_k)\big)
-\frac12 (\beta \epsilon_k)
\right] \notag \\
& \label{boundary theory cal}
\ \ \ \ \ +  N\int \frac{d^d k}{(2\pi)^d}
\frac{\beta}{2}\left[
\widetilde C(\beta,\ell)-\widetilde C(\infty,\ell)
\right],
\end{align}
where 
\begin{equation*}
    \tilde{C}(\beta, \ell)=-\frac{1}{\beta}\log(\ell) + A(\beta, \ell),
\end{equation*}
to ensure $\ell \rightarrow 1/\ell$ inversion symmetry. The function $A(\beta, \ell)$ can be any function satisfying $A(\beta, \ell) = A(\beta, 1/\ell)$. One of the simplest choice is that $A(\beta, \ell)$ is a function of $\beta$ only.


Now, we claim that the two quantities amazingly coincide each other when the mass gap in large N mean field theory of O(N) vector model vanishes. This limit can be achieved either by not to impose any constraint on the theory or imposing the $\delta(\tilde n_\ell \tilde n_\ell-N)$ constraint but becoming gapless at zero coupling limit together with $\ell=1$.

In the case of $m$=0, by an identification that $\ell=\ell^{(H)}$ and $ \beta =2t$, which leads $\beta \epsilon_{\vec{k}} = 2|k|t $,
we can derive the following correspondence:
\begin{equation}
\label{CoRResponDence}
\boxed{   \beta \bigg[ 
{\mathcal{F}(\ell,\beta)-\mathcal{F}(\ell,\infty)}
\bigg]\bigg|^{\ell=\ell^{(H)},\beta=2t} = s(\ell^{(H)},t)  - s(\ell^{(H)}, \infty) }
\end{equation}
where we have chosen that
\begin{align}
\label{diff. of const.}
    A(\beta, \ell) - A(\infty, \ell) =  \frac{\pi i}{\beta}.
\end{align}

\subsection{Gap equation} 
We note that in the previous section, the correspondence(\ref{CoRResponDence}) is made only when the mass gap vanishes. In the case that we do not impose any constraint on the system, this correspondence is hold. However, another case when we impose the $\delta$ function constraint, the correspondence holds only when $\ell=1$ and the zero coupling limit. We need to discuss this point in this subsection.
The equation of motion for the auxiliary field $\lambda$ has already been derived in \eqref{gap eq}.
Therefore, we begin directly with the large $N$ gap equation for our discussion. 
A detailed derivation of the momentum space form of \eqref{gap eq} is provided in Appendix A.

\paragraph{Evaluation of gap equation} 
For $\ell=1$, the covariant derivative $D_\tau$ reduces to the ordinary derivative $\partial_\tau$ and the shifted Matsubara frequency $\tilde{\omega}_n$ reduces to the Matsubara frequency $\omega_n$. 
Hence, in momentum space, \eqref{gap eq} takes the form
\begin{align}
\label{gap eq in momentum space}
\frac{1}{g}
= \frac{1}{\beta}\sum_{\omega_n} \int\frac{d^dk}{(2\pi)^d}
\frac{1}{\omega_n^2+\epsilon_{\vec{k}}^{2}},
\end{align}
where $\epsilon_{\vec{k}}^{2}=\vec{k}^2+m^2.$
Using the result for the Matsubara frequency summation \eqref{Matsubara sum in BE} is
\begin{align}
\label{Using the result for the Matsubara frequency summation and gap eq in momentum space }
\frac{1}{g}
=\int\frac{d^dk}{(2\pi)^d} \frac{1}{2\epsilon_{\vec{k}}}
\bigg[1+\frac{2}{e^{\beta\epsilon_{\vec{k}}}-1}\bigg].
\end{align}
Set the $d$-dimensional measure in spherical coordinates
\begin{align}
\label{d dimensional measure in spherical coordinates}
d^dk = k^{d-1}dk\, d\Omega_{d-1}, 
\quad
\Omega_{d-1}=\frac{2\pi^{d/2}}{\Gamma(d/2)},
\end{align}
where $k=|\vec{k}|$. We use \eqref{d dimensional measure in spherical coordinates} and obtain gap equation is
\begin{align}
\label{gap eq sol in momentum space}
\frac{1}{g}
&=
\frac{\Omega_{d-1}}{(2\pi)^d}
\int_{0}^{\Lambda} dk \frac{k^{d-1}}{2\sqrt{{k}^2+m^2}}
\bigg[1+\frac{2}{e^{\beta \sqrt{k^2+m^2}}-1} \bigg].
\end{align}
Instead using $\epsilon_{\vec{k}}^{2}=\vec{k}^2+m^2$, we change the integration variable from $k$ to $\epsilon_{\vec{k}}$ to simplify the calculation.
\begin{equation}
\label{change of variable}
d\epsilon_{\vec{k}}
=\frac{k}{\sqrt{k^2+m^2}} dk
=\frac{k}{\epsilon_{\vec{k}}} dk.
\end{equation}
Substituting this relation \eqref{change of variable} into the gap equation, equivalently rewrite \eqref{gap eq sol in momentum space}
\begin{align}
\frac{1}{g}
&=
\frac{\Omega_{d-1}}{(2\pi)^d}
\int_{|m|}^{\tilde\Lambda} d\epsilon_{\vec{k}} \frac{(\epsilon_{\vec{k}}^2-m^2)^{\frac{d-2}{2}}}{2}
\bigg[1+\frac{2}{e^{\beta \epsilon_{\vec{k}}}-1}  \bigg],
\end{align}
where $\tilde\Lambda=\sqrt{\Lambda^2+m^2}$.
\paragraph{Evaluation of gap equation at $\ell$=1, d=2 and m $\neq$ 0}
Since in $d=2$, integration measure becomes
$d^2k = k dk d\Omega_1 \to d^2k = \epsilon_{\vec{k}}\, d\epsilon_{\vec{k}} d\Omega_1$.
Substituting this relation into the gap equation \eqref{Using the result for the Matsubara frequency summation and gap eq in momentum space }, we obtain
\begin{align}
\label{gap equation at undeformed, d=2 and massive}
\frac{1}{g}  
=\frac{1}{2\pi} \int_{|m|}^{\tilde\Lambda} d\epsilon_{\vec{k}}
\bigg[\frac{1}{2} +\frac{1}{e^{\beta\epsilon_{\vec{k}}}-1} \bigg],
\end{align}
where $\tilde\Lambda=\sqrt{\Lambda^2+m^2}$.
Substituting $y=e^{\beta\epsilon_{\vec{k}}}$ to evaluate \eqref{gap equation at undeformed, d=2 and massive}, we get the gap equation at $\ell=1$, $d=2$ and $m \neq 0$:
\begin{align}
\label{gap equation d=2 }
\frac{1}{g}
=\frac{1}{2\pi}
\bigg[ \frac{1}{2}(\tilde\Lambda-|m|)+\frac{1}{\beta}
\log \left(\frac{1-e^{-\beta\tilde\Lambda}}{1-e^{-\beta |m|}}  \right)\bigg].
\end{align}
To simplify this, using the identity
\begin{align}
\label{spcial case : identity for sinh}
\log\bigg(\sinh\left(\frac{\beta x}{2}\right) \bigg)
=\frac{\beta x}{2}+\log(1-e^{-\beta x})-\log 2.
\end{align}
Substituting $x=\tilde{\Lambda}$ and $x=|m|$ into \eqref{spcial case : identity for sinh}, the terms in brackets on the right hand side of \eqref{gap equation d=2 } can be combined as
\begin{align}
\log\left(\frac{\sinh\left(\beta\tilde\Lambda/2 \right) }{\sinh\left(\beta |m|/2\right)}    \right)
=\frac{\beta}{2}(\tilde\Lambda-|m|)+
\log\left(\frac{1-e^{-\beta\tilde\Lambda}}{1-e^{-\beta |m|}}   \right).
\end{align}
Then \eqref{gap equation d=2 } can be written compactly as
\begin{align}
\label{gap eq in compact form}
\frac{2\pi\beta}{g}
=\log\left(   \frac{\sinh\left(\beta\tilde\Lambda/2\right)}{\sinh\left(\beta |m|/2\right)}    \right).
\end{align}
Rearranging \eqref{gap eq in compact form} for $m$, we obtain
\begin{align}
\label{mass gap sol with large cutoff}
|m|=\frac{2}{\beta}
\operatorname{arcsinh} \left( e^{-2\pi\beta/g}\sinh\bigg(\frac{\beta\tilde\Lambda}{2}\bigg)
\right).
\end{align}
\paragraph{In large cutoff limit} $\Lambda\gg \beta^{-1},m$
\begin{align}
\sinh\left(\frac{\beta \tilde\Lambda}{2}\right)
\simeq \frac{1}{2}e^{\beta \Lambda/2},
\end{align}
where $\tilde\Lambda\simeq \Lambda$ and $1/g_c \equiv \Lambda/(4\pi)$. We can rewrite \eqref{mass gap sol with large cutoff} as
\begin{align}
\label{gap eq large cutoff limit with undeformed}
|m|
&\simeq \frac{2}{\beta}
\operatorname{arcsinh}\left(
\frac{1}{2} e^{-2 \pi \beta \left[\frac{1}{g}-\frac{1}{g_c} \right]  }    \right).
\end{align}
In the weak-coupling regime $0< g \ll g_c$, taking the limit $g \to 0^+$ forces $m = 0 $.

\paragraph{Evaluation of gap equation at $\ell$ $\neq$ 1, d=2 and m $\neq$ 0}
In this case, the background gauge field shifts only the Matsubara frequency through the covariant derivative, $\omega_n\to\widetilde{\omega}_n$. 
Since neither the spatial kinetic term nor the mass term is modified, the dispersion relation $\epsilon_{\vec{k}}=\sqrt{\vec{k}^{\,2}+m^2}$ remains unchanged, while the thermal distribution function is deformed, a modification known as the $\ell$-deformation.
We recall \eqref{Calculation of Matsubara sum by deformed}
\begin{align}
\frac{1}{\beta}\sum_n \frac{1}{\tilde\omega_n^2+\epsilon_{\vec{k}}^2}
=\frac{1}{2\epsilon_{\vec{k}}}
\left[\frac{\ell e^{-\beta\epsilon_{\vec{k}}}}{1-\ell e^{-\beta\epsilon_{\vec{k}}}}
-\frac{\ell e^{\beta\epsilon_{\vec{k}}}}{1-\ell e^{\beta\epsilon_{\vec{k}}}}
  \right].
\end{align}
Insert this into the gap equation \eqref{gap eq in momentum space} in $d=2$
\begin{align}
\frac{1}{g}
&=
\int_{0}^{\Lambda} \frac{d^2k}{(2\pi)^2}
\frac{1}{2\epsilon_{\vec{k}}}
\left[\frac{\ell e^{-\beta\epsilon_{\vec{k}}}}{1-\ell e^{-\beta\epsilon_{\vec{k}}}} -\frac{\ell e^{\beta\epsilon_{\vec{k}}}}{1-\ell e^{\beta\epsilon_{\vec{k}}}}     \right].
\end{align}
Changing variables from $k$ to $\epsilon_{\vec{k}}$, we get
\begin{align}
\label{deformed gap eq in d=2 }
\frac{1}{g}
&=
\frac{1}{4\pi}
\int_{|m|}^{\tilde\Lambda}
d\epsilon_{\vec{k}}
\left[
\frac{\ell e^{-\beta\epsilon_{\vec{k}}}}{1-\ell e^{-\beta\epsilon_{\vec{k}}}}
-\frac{\ell e^{\beta\epsilon_{\vec{k}}}}{1-\ell e^{\beta\epsilon_{\vec{k}}}}
\right].
\end{align}
To evaluate \eqref{deformed gap eq in d=2 }, we rewrite each term in the integrand as a logarithmic derivative with respect to $\epsilon_{\vec{k}}$:
\begin{align}
\frac{d}{d\epsilon_{\vec{k}}}\log(1-\ell e^{-\beta\epsilon_{\vec{k}}})
&=\beta\frac{\ell e^{-\beta\epsilon_{\vec{k}}}}
{1-\ell e^{-\beta\epsilon_{\vec{k}}}},
\\
\frac{d}{d\epsilon_{\vec{k}}}\log(1-\ell e^{\beta\epsilon_{\vec{k}}})
&=-\beta\frac{\ell e^{\beta\epsilon_{\vec{k}}}}
{1-\ell e^{\beta\epsilon_{\vec{k}}}}.
\end{align}
Therefore, the integral becomes
\begin{align}
\label{gap eq sol with deformed}
\frac{1}{g}
&=
\frac{1}{4\pi\beta}
\left[ \log\left(\frac{1-\ell e^{-\beta\tilde\Lambda}}{1-\ell e^{-\beta |m|}}\right)
+ \log\left(\frac{1-\ell e^{\beta\tilde\Lambda}}{1-\ell e^{\beta |m|}}\right) \right].
\end{align}
As a consistency check, setting $\ell=1$ in \eqref{gap eq sol with deformed} gives
\begin{align}
\frac{1}{g}
&=\frac{1}{4\pi\beta}
\log \left(
\frac{2-2\cosh\left(\beta\widetilde\Lambda\right) }{2-2\cosh\left(\beta |m|\right) }
\right).
\end{align}
To simplify this, using the identity
\begin{equation}
2-2\cosh x=-4\sinh^2 \left(\frac{x}{2}\right),
\end{equation}
we get
\begin{align}
\frac{1}{g}=\frac{1}{2\pi\beta}\log \left(
\frac{\sinh\left(\beta\widetilde\Lambda/2\right)}{\sinh\left(\beta |m| /2\right)}
\right).
\end{align}
This is recovered in \eqref{gap eq in compact form} at $\ell=1$. 
\paragraph{In large cutoff limit} $\Lambda\gg \beta^{-1},m$ \eqref{gap eq sol with deformed} simplifies to
\begin{align}
\frac{1}{g}
&=
\frac{1}{4\pi\beta}\log\ell
+ \frac{1}{g_c} - \frac{1}{4\pi\beta} \log(2\ell\cosh(\beta |m|)-\ell^2-1),
\end{align}
where $1/g_c \equiv \Lambda/(4\pi)$.
\begin{align}
\frac{1}{g} - \frac{1}{g_c} - \frac{1}{4\pi\beta}\log\ell
= - \frac{1}{4\pi\beta} \log(2\ell\cosh(\beta |m|)-\ell^2-1).
\end{align}
Multiplying by $-4\pi\beta$ and exponentiating, we obtain 
\begin{align}
\label{gap equation ell deformed}
\ell  e^{-4\pi\beta\left[\frac{1}{g} - \frac{1}{g_c}\right]} + \ell^2 +1
= 2\ell\cosh(\beta |m|).
\end{align}
Take the weak-coupling $0<g\ll g_c$ limit with $g\to 0^{+}$. 
Then, $\exp[-4\pi\beta \big[\frac{1}{g} - \frac{1}{g_c}\big]] \to 0$, reduces to
\begin{align}
\label{ell sol in two branch}
\ell+\frac{1}{\ell} = e^{\beta |m|}+e^{-\beta |m|}.
\end{align}
The $\ell$ solution is
\begin{align}
\label{small coupling limit}
\beta |m|
= 
\begin{cases}
\log\ell 
& \text{for  } \ell \geq 1, \\
-\log\ell 
& \text{for } 0< \ell \leq 1.
\end{cases}
\end{align}
Since the mass $m$ enters the dispersion relation through $m^{2}$, the minimum excitation energy is $\epsilon_{\vec{0}}=|m|$. 
For $\ell\neq1$, a nonzero mass gap is generated.

But, the undeformed weak-coupling result obtained by taking $g\to0^+$ in \eqref{gap eq large cutoff limit with undeformed} is recovered at $\ell=1$ as
\begin{align}
\beta m = 0 \qquad \rightarrow \qquad m=0.
\end{align}

\section*{Acknowledgement}
J.H.O thanks his W.J. and Y.J. J.H.O also thanks to Prof. Ki-seok Kim at POSTECH for useful discussion and the discussion was very much helpful to set up our business in earlier period. He also thanks CQUeST for hospitality.

\begin{appendices}
\newpage
\section{O(N) non-linear sigma model in 1+2 dim at large N limit}
\label{Appendix A : O(N) non-linear sigma model in 2+1 dim case at large N limit}
In this appendix, we provide a detailed derivation of the large N effective action for the O(N) non-linear sigma model discussed in Section~\ref{On Calculation of Free Energy in CFT}. 
We recall that, with the background gauge field set to zero, the partition function of the O(N) non-linear sigma model is given by~\eqref{The action for O(N) vector model in large N limit}
\begin{align}
    & Z = \int [D\Tilde{n_{l}}][D \lambda] \exp \bigg[ -\frac{1}{2g}\int d^{2} r \int_{0}^{\beta} d\tau [ (\nabla_{r}\tilde{n}_{l})^{2} +  (\partial_{\tau} \tilde{n}_{l})^{2} + i\lambda (\tilde{n}^{2}_{l} -N)   ]                            \bigg] \notag \\
    &= \int [D\lambda]   \det\left| \frac{1}{g}(-\partial_\tau^2 - \nabla^2 + i\lambda) \right|^{-N/2} 
    \exp\left\{ -\frac{1}{2g} \int d^2r \int_0^\beta d\tau (-i\lambda N) \right\} \notag \\
    &= \int [D\lambda] \exp \left\{
    -\frac{N}{2}   \mathrm{tr} \ln \left| -\partial_\tau^2 - \nabla^2 + i\lambda \right| 
    + \frac{N}{2}   \mathrm{tr} \ln g 
    + \frac{iN}{2g} \int d^2r \int_0^\beta d\tau   \lambda 
    \right\}.
\end{align}
In the second line, integration by parts is performed and the Gaussian functional integral over the $N$-component field $\widetilde{n}$ is evaluated. 
Since each component contributes a factor $(\det \hat{A} )^{-1/2}$, the integration over all $N$ components gives $(\det \hat{A} )^{-N/2}$. 
In this case, the operator $\hat{A}$ is given by $( -\partial_\tau^2 - \nabla^2 + i\lambda) $. 
In the third line, using the identity $\det \hat{A} =\exp(\mathrm{tr} \ln \hat{A} )$, the determinant is rewritten in the trace-log form.
The partition function can be written as $Z=\int[D\lambda]\exp\!\left[-N S_{\mathrm{eff}}[\lambda]\right]$, where $S_{\mathrm{eff}}[\lambda]$ is the effective action for the auxiliary field $\lambda$. The saddle-point condition is
\begin{align}
\frac{\delta S_{\mathrm{eff}}[\lambda]}{\delta \lambda(\vec{r '}, \tau ')} =0.
\end{align}
The variation of the effective action with respect to the auxiliary field $\lambda$ is given by
\begin{align}
\label{appendixA : The variation of the effective action}
\frac{\delta S_{\text{eff}}[\lambda]}{\delta \lambda(\vec{r'},\tau ')} 
&= \frac{\delta}{\delta \lambda(\vec{r'},\tau')} \Bigg[ 
        \frac{1}{2}   \mathrm{tr} \ln \big| -\partial_{\tau}^{2} - \nabla^2 + i\lambda(\vec{r},\tau) \big|
    - 
        \int d^2r \int_0^\beta d\tau \frac{i \lambda(\vec{r},\tau)}{2g} 
\Bigg].
\end{align}
To evaluate the first term in \eqref{appendixA : The variation of the effective action}, we use the trace-log variation 
\begin{align}
\label{appendixA : trace-log relation}
&\delta   \mathrm{tr} \log \hat{A} = \sum_n \frac{1}{\lambda_n} \delta \lambda_n = \mathrm{tr}\big( \hat{A}^{-1} \delta \hat{A} \big).
\end{align}
Applying \eqref{appendixA : trace-log relation} to
\eqref{appendixA : The variation of the effective action}, we obtain
\begin{align}
\frac{\delta S_{\mathrm{eff}}[\lambda]}
{\delta\lambda(\vec{r '},\tau')}
&=
\frac{i}{2}
\int d^2r\int_0^\beta d\tau
\left \langle 
\vec r,\tau \left| \hat A^{-1} \right| \vec r,\tau
\right\rangle
\delta^{(2)}(\vec r-\vec{r '}) \delta(\tau-\tau')
-\frac{i}{2g}
\int d^2r\int_0^\beta d \tau
\delta^{(2)}(\vec r-\vec{r '}) \delta(\tau-\tau')
\notag\\
&=
\frac{i}{2}
\int d^2r\int_0^\beta d\tau
\left[
\left\langle
\vec r,\tau \left| \hat A^{-1} \right| \vec r,\tau
\right\rangle 
-\frac{1}{g} \right] \delta^{(2)}(\vec r-\vec{r '}) \delta(\tau-\tau').
\label{appendixA : saddle-point equation}
\end{align}
This corresponds to the result in Ref.\cite{5},
\begin{align}
\label{appendixA : large N gap equation}
\frac{1}{g}
=
\left\langle
\vec{r },\tau \left| \frac{1}{-\partial_\tau^2-\nabla^2+i\lambda(\vec r, \tau)}\right| \vec{r },\tau
\right\rangle.
\end{align}
This is the large $N$ gap equation in position space.
It shows that  $i \langle \lambda \rangle$  plays the role of the mass squared $m^2$ and parametrizes the saddle-point value of $\lambda$.  
Alternatively, the gap equation can be considered by first specifying
the saddle-point value of the auxiliary field.
At the saddle-point, we write $i\langle\lambda\rangle=C$, where $C$ is independent of $\vec{r}$ and $\tau$. 
In momentum space, the inverse quadratic kernel is $(\Hat{\widetilde{A}})^{-1}(\omega_n,\vec{k}) =1/(\omega_n^2+\vec{k}^{2}+C)$.
The tilde denotes the momentum space representation of the position space operator $\Hat{A}$. 
Since all terms in $\Hat{\widetilde{A}}$ have mass dimension two, writing $C$ as $m^2$ gives $i\langle\lambda\rangle=m^2$.
The corresponding dispersion relation $\epsilon_{\vec{k}}=\sqrt{\vec{k}^{2}+m^2}$ identifies $m$ as the mass associated with the saddle-point value of the auxiliary field.
Substituting $i\langle\lambda\rangle=m^2$ into \eqref{appendixA : The variation of the effective action} gives
\begin{equation}
S_{\mathrm{eff}}(m^2)
=
\frac{1}{2}\mathrm{tr} \ln \left(-\partial_\tau^2-\nabla^2+m^2 \right) -\int d^2r\int_0^\beta d\tau \frac{m^2}{2g}.
\end{equation}
Similarly, the saddle-point equation for $m^2$ is
$\partial S_{\mathrm{eff}}(m^2)/\partial m^2=0$, which gives
\begin{align}
\frac{1}{g}
=\frac{1}{\beta V_2}\mathrm{tr}\left( \frac{1}{-\partial_\tau^2-\nabla^2+m^2}
\right),
\end{align}
where $V_2 = \int d^2 r$ is the spatial volume and $\beta = \int_0^{\beta} d \tau$.  
Equation~\eqref{appendixA : large N gap equation} is recovered.
We recall the definition of the partition function and free energy as
\begin{align}
&  Z[\lambda] = \int [D\lambda] \exp( -NS_{\text{eff}}[\lambda]  ) \xrightarrow[]{ N \to \infty } \int [D\lambda] \exp( -NS_{\text{eff}}(i\lambda=m^2)  ), \ \
 F = -\frac{1}{\beta} \log Z,
\end{align}
where Boltzmann constant is set to $k_B=1$.
Thus, the effective action and free energy are
\begin{align}
&S_{\text{eff}}(m^2) = \frac{1}{2}   \mathrm{tr} \log \left( -\partial_{\tau}^{2} - \nabla^{2} + m^{2} \right)
- \frac{1}{2}   \mathrm{tr} \log g 
- \int d^{2}r \int_{0}^{\beta} d\tau   \frac{m^{2}}{2g},  
\\ & F = +\frac{1}{\beta} N S_{\text{eff}}(m^2)
= \frac{N}{\beta} \left[
\frac{1}{2}   \mathrm{tr} \log \left( -\partial_{\tau}^{2} - \nabla^{2} + m^{2} \right)
- \int d^{2}r \int_{0}^{\beta} d\tau   \frac{m^{2}}{2g}
\right]. \label{free energy before evaluation 2}
\end{align}
The term $-\frac{1}{2}\mathrm{tr}\log g$ is independent of the auxiliary field $\lambda$ and contains no dynamical information.
Since it contributes only an additive constant, it cancels out when free energy differences are taken.

Having obtained the effective action and the free energy, we now evaluate in momentum space.
To evaluate the free energy in \eqref{free energy before evaluation}, we consider its two contributions separately.
We first express the trace-log term in momentum space as
\begin{align}
&\int d^{2}r \int_{0}^{\beta} d\tau 
\int \frac{d\omega   d\omega'}{(2\pi)^{2}} 
\int \frac{d^{2}\vec{k}   d^{2}\vec{k}'}{(2\pi)^{4}}  
[ \langle \vec{r}, \tau | \vec{k}, \omega \rangle 
\langle \vec{k}, \omega | \ln \big( -\partial_{\tau'}^{2} - \nabla^{2} + m^{2} \big) | \vec{k}', \omega' \rangle 
\langle \vec{k}', \omega' | \vec{r}, \tau \rangle ] \notag \\
&= \int d^{2}r \int_{0}^{\beta} d\tau 
\int \frac{d\omega   d\omega'}{(2\pi)^{2}} 
\int \frac{d^{2}\vec{k}   d^{2}\vec{k}'}{(2\pi)^{4}}  
e^{i (\vec{k} \cdot \vec{r} - \omega \tau)} 
e^{-i (\vec{k}' \cdot \vec{r} - \omega' \tau)} 
\ln \left( \vec{k}'^{ 2} + \omega'^{2} + m^{2} \right) 
  \delta^{(2)}(\vec{k} - \vec{k}') \delta(\omega - \omega' ) \notag \\
&= \int d^{2}r \int_{0}^{\beta} d\tau 
\int \frac{d\omega}{2\pi} 
\int \frac{d^{2}\vec{k}}{(2\pi)^{2}}  
\ln \left( \vec{k}^{ 2} + \omega^{2} + m^{2} \right).
\end{align}
In order to account for finite-temperature effects, we replace the continuous frequency integral by a Matsubara frequency summation, $\omega \rightarrow \omega_n=\frac{2\pi n}{\beta}, \ \ \int_{-\infty}^{\infty}\frac{d\omega}{2\pi} \rightarrow \frac{1}{\beta}\sum_{n=-\infty}^{\infty}$,  leading to
\begin{align}
   \int d^{2}r \int_{0}^{\beta} d\tau  \frac{1}{\beta} \sum_{\omega_{n}} \int \frac{d^{2}\vec{k}}{(2\pi)^{2}} \ln \left( \vec{k}^{ 2} + \omega_{n}^{2} + m^{2} \right).
\end{align}
The free energy in momentum space is given by
\begin{align}
F &= N V_2
\bigg[ 
\frac{1}{2\beta} \sum_{\omega_n} \int \frac{d^{2}\vec{k}}{(2\pi)^2} 
\ln \big( \vec{k}^{ 2} + \omega_n^2 + m^2 \big) 
- \frac{m^2}{2g}
\bigg],
\end{align}
where $\int d^{2}r \int_{0}^{\beta} d\tau$=$\beta V_2, \  V_{2}$ called system spatial volume in Ref.\cite{4}.
Dividing by $V_2$, we obtain the free energy density as
\begin{equation}
\label{We get the Free energy density}
    \frac{F}{V_2} \equiv \mathcal{F} = \frac{N}{2\beta} \sum_{\omega_n} \int \frac{d^{2}\vec{k}}{(2\pi)^2} 
\ln \big( \vec{k}^{ 2} + \omega_n^2 + m^2 \big) 
- \frac{N m^2}{2g}. 
\end{equation}

\section{Contour Integral Evaluation of $\ell$-deformed Matsubara Frequency Summation}
\label{Appendix B : More detail on the Calculation of Free Energy in CFT}
\paragraph{$\ell = 1$ : Undeformed Matsubara frequency summation}
\begin{align}
\label{pre Undeformed Matsubara frequency summation}
\frac{1}{\beta} \sum_n \frac{1}{\omega_n^2 + \epsilon_{\vec{k}}^2} 
&=  \oint_C \frac{dz}{2\pi i} \frac{1}{-z^2 + \epsilon_{\vec{k}}^2}  \frac{1}{e^{\beta z} - 1}  \notag \\
&= \oint_{C^{\prime}} \frac{dz}{2\pi i} (-1) \left[ \frac{1}{z - \epsilon_{\vec{k}}} - \frac{1}{z + \epsilon_{\vec{k}}} \right]  \frac{1}{2\epsilon_{\vec{k}}}  \frac{1}{e^{\beta z} - 1},
\end{align}
where $\beta$ denotes the inverse temperature and Boltzmann constant is set to $k_B=1$.
For $\ell=1$, the covariant derivative $D_\tau$ reduces to the ordinary derivative $\partial_\tau$ and the shifted Matsubara frequency $\tilde{\omega}_n$ reduces to the Matsubara frequency $\omega_n$. 
The Matsubara frequency summation is written as a contour integral using the Bose-Einstein distribution function.
We deform the contour and pick $z_0=\pm\epsilon_{\vec{k}}$ as the poles.
Then, \eqref{pre Undeformed Matsubara frequency summation} becomes
\begin{align}
\label{undeformed Matsubara frequency summation}
\frac{1}{\beta} \sum_n \frac{1}{\omega_n^2 + \epsilon_{\vec{k}}^2} &= \frac{1}{2\epsilon_{\vec{k}}} \left[ 1 + \frac{2}{e^{\beta \epsilon_{\vec{k}}} - 1} \right]
= \frac{1}{2\epsilon_{\vec{k}}} \left[ \frac{e^{-\beta \epsilon_{\vec{k}}}}{1-e^{-\beta \epsilon_{\vec{k}}}} - \frac{e^{\beta \epsilon_{\vec{k}}}}{1-e^{\beta \epsilon_{\vec{k}}}} \right].
\end{align}
Multiplying by $2\epsilon_{\vec{k}}$, we obtain
\begin{align}
 \frac{1}{\beta} \sum_n \frac{2\epsilon_{\vec{k}}}{\omega_n^2 + \epsilon_{\vec{k}}^2} & \label{cal step : undeformed Matsubara frequency summation 1}= \left[ 1 + \frac{2}{e^{\beta \epsilon_{\vec{k}}} - 1} \right] \\
& \label{cal step : undeformed Matsubara frequency summation 2}=  \left[ \frac{e^{-\beta \epsilon_{\vec{k}}}}{1-e^{-\beta \epsilon_{\vec{k}}}} - \frac{e^{\beta \epsilon_{\vec{k}}}}{1-e^{\beta \epsilon_{\vec{k}}}} \right].
\end{align}
The above equations can be expressed in terms of derivatives with respect to $\epsilon_{\vec{k}}$, we obtain
\begin{align}
\label{cal step : undeformed Matsubara frequency summation 3}
\frac{1}{\beta} \sum_n \frac{\partial}{\partial \epsilon_{\vec{k}}} \log(\omega_n^2 + \epsilon_{\vec{k}}^2) 
& = \frac{1}{\beta} \frac{\partial}{\partial\epsilon_{\vec{k}}} \left[ \log(2 -2 \cosh(\beta \epsilon_{\vec{k}} )) + \log(-1) \right] \\
& \label{cal step : undeformed Matsubara frequency summation 4}= \frac{1}{\beta} \frac{\partial}{\partial\epsilon_{\vec{k}}} \left[ \log(2 -2 \cosh(\beta \epsilon_{\vec{k}} )) \right].
\end{align}
Integrating \eqref{cal step : undeformed Matsubara frequency summation 3} and \eqref{cal step : undeformed Matsubara frequency summation 4} with respect to $\epsilon_{\vec{k}}$, we get
\begin{align}
\frac{1}{\beta} \sum_n \log(\omega_n^2 + \epsilon_{\vec{k}}^2) 
& = \frac{1}{\beta}  \log(2 -2 \cosh(\beta \epsilon_{\vec{k}} )) + \frac{\log(-1)}{\beta} + C_1(\beta) \notag \\
&\label{ell=1 MFS results 1} = \frac{1}{\beta}  \log(2 -2 \cosh(\beta \epsilon_{\vec{k}} )) + \bar{C}(\beta) \\
& \label{ell=1 MFS results 2} = \frac{1}{\beta}  \log(2 -2 \cosh(\beta \epsilon_{\vec{k}} )) + C_2(\beta),
\end{align}
where $C_1(\beta) + \frac{\log(-1)}{\beta} = \bar{C}(\beta)$.
The constants $C_1(\beta), C_2(\beta)$ and $\bar{C}(\beta)$ are independent of $\epsilon_{\vec{k}}$.
\paragraph{\texorpdfstring{$\ell$}{ell}-deformation : Generalized Matsubara frequency summation}
In this case, the background gauge field shifts only the Matsubara frequency through the covariant derivative, $\omega_n\to\widetilde{\omega}_n$.
So, the Bose-Einstein distribution $(e^{\beta z}-1)^{-1}$ is correspondingly generalized to the $\ell$-deformed distribution $(\ell^{-1}e^{\beta z}-1)^{-1}$.
The $\ell$-deformed Matsubara frequency summation can be evaluated in the same way as in the $\ell=1$ case,
\begin{align}
\frac{1}{\beta} \sum_n \frac{1}{\tilde{\omega}_n^2 + \epsilon_{\vec{k}}^2}
&= \oint_C \frac{dz}{2\pi i} \frac{1}{-z^2 + \epsilon_{\vec{k}}^2}  \frac{1}{\ell^{-1} e^{\beta z} - 1} \notag \\
&= \oint_{C^{\prime}} \frac{dz}{2\pi i} (-1)\left[ \frac{1}{z - \epsilon_{\vec{k}}} - \frac{1}{z + \epsilon_{\vec{k}}} \right]  \frac{1}{2\epsilon_{\vec{k}}}  \frac{1}{\ell^{-1} e^{\beta z} - 1}.
\end{align}
Then,
\begin{align}
\label{ell deformed Matsubara frequency summation}
\frac{1}{\beta} \sum_n \frac{1}{\tilde{\omega}_n^2 + \epsilon_{\vec{k}}^2}
& = \frac{1}{2\epsilon_{\vec{k}}} \left[ \frac{\ell e^{-\beta \epsilon_{\vec{k}}}}{1 - \ell e^{-\beta \epsilon_{\vec{k}}}} - \frac{\ell e^{\beta \epsilon_{\vec{k}}}}{1 - \ell e^{\beta \epsilon_{\vec{k}}}} \right].
\end{align} 
We multiply both sides by $2\epsilon_{\vec{k}}$. 
The resulting can be expressed in terms of derivatives with respect to $\epsilon_{\vec{k}}$ as
\begin{align}
\label{cal step : ell deformed Matsubara frequency summation}
\frac{1}{\beta} \sum_n \frac{\partial}{\partial \epsilon_{\vec{k}}} \log(\tilde{\omega}_n^2 + \epsilon_{\vec{k}}^2)
&= \frac{1}{\beta} \frac{\partial}{\partial\epsilon_{\vec{k}}} \left[ \log(1 - \ell e^{-\beta \epsilon_{\vec{k}}}) + \log(1 - \ell e^{\beta \epsilon_{\vec{k}}}) \right].
\end{align}
Integrating \eqref{cal step : ell deformed Matsubara frequency summation} with respect to $\epsilon_{\vec{k}}$, one finds
\begin{equation}
\label{cal step : ell deformed Matsubara frequency summation 2}
    \frac{1}{\beta} \sum_n \log(\tilde{\omega}_n^2 + \epsilon_{\vec{k}}^2)
 = \frac{1}{\beta}  \log\left( \ell^2 + 1 - 2\ell \cosh(\beta \epsilon_{\vec{k}}) \right) +C_3(\beta, \ell).
\end{equation}
\paragraph{Consistency check at $\ell$=1}
For $\ell = 1$, the result of \eqref{cal step : ell deformed Matsubara frequency summation 2} is
\begin{align}
\label{cal step : Consistency check at ell=1}
\frac{1}{\beta} \sum_n \log(\tilde{\omega}_n^2 + \epsilon_{\vec{k}}^2)
& = \frac{1}{\beta} \log\left( 2 - 2\cosh(\beta \epsilon_{\vec{k}}) \right) +C_3(\beta, \ell=1).
\end{align}
For comparison, the results of equations \eqref{ell=1 MFS results 1} and \eqref{ell=1 MFS results 2} are
\begin{align}
\label{cal step : Consistency check at ell=1, 1}
\frac{1}{\beta} \sum_n \log(\tilde{\omega}_n^2 + \epsilon_{\vec{k}}^2)
&=  \frac{1}{\beta}  \log(2 -2 \cosh(\beta \epsilon_{\vec{k}} ))  + \bar{C}(\beta, \ell=1) \\
&\label{cal step : Consistency check at ell=1, 2}= \frac{1}{\beta}   \log(2 -2 \cosh(\beta \epsilon_{\vec{k}} )) + C_2(\beta, \ell=1).
\end{align}
The $\epsilon_{\vec{k}}$ dependent parts of \eqref{cal step : Consistency check at ell=1} $\sim$ \eqref{cal step : Consistency check at ell=1, 2} are identical. 
Therefore, identifying the integration constants as
$C_3(\beta,\ell=1)=\bar{C}(\beta,\ell=1)=C_2(\beta,\ell=1)$
recovers the undeformed result.

\paragraph{Structure of Poles}
The poles of the $\ell$-deformed are determined by $\ell^{-1}e^{\beta z_n}-1=0.$
Solving this condition gives 
\begin{equation}
z_n=\frac{1}{\beta}\left(\log\ell + 2\pi i n\right) \text{  with $n\in\mathbb{Z}$}.
\end{equation}
Introducing the shifted Matsubara frequency, $z_n \equiv -i\widetilde{\omega}_n$,
\begin{equation}
\widetilde{\omega}_n
= \omega_n + \frac{i}{\beta}\log\ell
\text{ with $\omega_n= \frac{2\pi n}{\beta}$}.
\end{equation}
Return to \eqref{cal step : ell deformed Matsubara frequency summation 2} and evaluate the logarithmic Matsubara frequency summation is
\begin{equation}
\label{cal step : ell deformed Matsubara frequency summation 3}
    \frac{1}{\beta} \sum_n \log(\tilde{\omega}_n^2 + \epsilon_{\vec{k}}^2)
= \frac{1}{\beta} \sum_n \left[ \log(i\tilde{\omega}_n + \epsilon_{\vec{k}}) + \log(-i\tilde{\omega}_n + \epsilon_{\vec{k}}) \right].
\end{equation}
We evaluate the contribution from the first logarithmic term in \eqref{cal step : ell deformed Matsubara frequency summation 3}. Differentiating it with respect to $\epsilon_{\vec{k}}$, we obtain
\begin{align}
\label{cal step : ell deformed Matsubara frequency summation 4}
\frac{1}{\beta} \sum_n \frac{\partial}{\partial\epsilon_{\vec{k}}}  \log(i\tilde{\omega}_n + \epsilon_{\vec{k}})  & =
\frac{1}{\beta} \sum_n \frac{1}{i\tilde{\omega}_n + \epsilon_{\vec{k}}}.
\end{align}
The equation \eqref{cal step : ell deformed Matsubara frequency summation 4} can be expressed as a contour integral
\begin{align}
\frac{1}{\beta} \sum_n \frac{1}{i\tilde{\omega}_n + \epsilon_{\vec{k}}}
& = \oint_C \frac{dz}{2 \pi i} (-1) \frac{1}{z - \epsilon_{\vec{k}}}  \frac{1}{\ell^{-1} e^{\beta z} - 1}.
\end{align}
The contour to enclose the pole at $z_0=\epsilon_{\vec{k}}$, we get
\begin{align}
 \oint_{C^{\prime}} \frac{dz}{2 \pi i} (-1) \frac{1}{z - \epsilon_{\vec{k}}}  \frac{1}{\ell^{-1} e^{\beta z} - 1}  & = - \Res[(-1)\frac{1}{z - \epsilon_{\vec{k}}}  \frac{1}{\ell^{-1} e^{\beta z} - 1}, z_{0}=\epsilon_{\vec{k}} ] \notag \\
 & \qquad \qquad + \int_{C_{\infty}} \frac{dz}{2 \pi i} (-1) \frac{1}{z -  \epsilon_{\vec{k}}}  \frac{1}{\ell^{-1} e^{\beta z} - 1}.
\end{align}
The residue at $z_0=\epsilon_{\vec{k}}$ and the contribution from the contour at infinity give
\begin{align}
\label{cal step : ell deformed Matsubara frequency summation 5}
\oint_{C^{\prime}} \frac{dz}{2\pi i} (-1) \frac{1}{z - \epsilon_{\vec{k}}} \frac{1}{\ell^{-1} e^{\beta z} - 1}
&= \frac{1}{\ell^{-1} e^{\beta \epsilon_{\vec{k}}} - 1} + \frac{1}{2}.
\end{align}
Substituting \eqref{cal step : ell deformed Matsubara frequency summation 5} into \eqref{cal step : ell deformed Matsubara frequency summation 4} and integrating with respect to $\epsilon_{\vec{k}}$, we get
\begin{align}
\label{cal step : ell deformed Matsubara frequency summation 8}
\frac{1}{\beta} \sum_n \log(i\tilde{\omega}_n + \epsilon_{\vec{k}})
&  = \frac{1}{\beta} \log\left(1-\ell e^{-\beta \epsilon_{\vec{k}}} \right) + \frac{1}{2} \epsilon_{\vec{k}} + \tilde{C}_{1}(\beta, \ell).
\end{align}
We evaluate the contribution from the second logarithmic term in \eqref{cal step : ell deformed Matsubara frequency summation 3}.
Differentiating it with respect to $\epsilon_{\vec{k}}$, one finds
\begin{align}
\label{cal step : ell deformed Matsubara frequency summation 6}
\frac{1}{\beta} \sum_n \frac{\partial}{\partial\epsilon_{\vec{k}}} \log(-i\tilde{\omega}_n + \epsilon_{\vec{k}}) 
&= \frac{1}{\beta} \sum_n \frac{1}{-i\tilde{\omega}_n + \epsilon_{\vec{k}}}.
\end{align}
The equation \eqref{cal step : ell deformed Matsubara frequency summation 6} can be expressed as a contour integral
\begin{align}
\frac{1}{\beta} \sum_n \frac{1}{-i\tilde{\omega}_n + \epsilon_{\vec{k}}}
&= \oint_C \frac{dz}{2 \pi i} \frac{1}{z + \epsilon_{\vec{k}}} \frac{1}{ \ell^{-1} e^{\beta z} - 1}.
\end{align}
Similarly, the contour to enclose the pole at $z_0= -\epsilon_{\vec{k}}$, we obtain
\begin{align}
\oint_{C^{\prime}} \frac{dz}{2 \pi i} \frac{1}{z + \epsilon_{\vec{k}}} \frac{1}{ \ell^{-1} e^{\beta z} - 1}     &  = -\Res[\frac{1}{z + \epsilon_{\vec{k}}} \frac{1}{ \ell^{-1} e^{\beta z} - 1} , z_{0}=-\epsilon_{\vec{k}} ]  \notag \\
 & \qquad \qquad \quad + \int_{C_{\infty}} \frac{dz}{2 \pi i} \frac{1}{z + \epsilon_{\vec{k}}}  \frac{1}{\ell^{-1} e^{\beta z} - 1}.
\end{align}
The residue at $z_0=-\epsilon_{\vec{k}}$ and the contribution from the contour at infinity give
\begin{align}
\label{cal step : ell deformed Matsubara frequency summation 7}
\oint_{C^{\prime}} \frac{dz}{2 \pi i}  \frac{1}{z + \epsilon_{\vec{k}}} \frac{1}{ \ell^{-1} e^{\beta z} - 1}   & = \frac{-1}{\ell^{-1} e^{-\beta \epsilon_{\vec{k}}} - 1} - \frac{1}{2}.
\end{align}
Substituting \eqref{cal step : ell deformed Matsubara frequency summation 7} into \eqref{cal step : ell deformed Matsubara frequency summation 6} and integrating with respect to $\epsilon_{\vec{k}}$, we get
\begin{align}
\label{cal step : ell deformed Matsubara frequency summation 9}
\frac{1}{\beta} \sum_n \log(-i\tilde{\omega}_n + \epsilon_{\vec{k}})
&= -\frac{1}{2} \epsilon_{\vec{k}} + \frac{1}{\beta} \log\left(1 - \ell e^{\beta \epsilon_{\vec{k}}} \right) + \tilde{C}_{2}(\beta, \ell).
\end{align}
Combining the results in \eqref{cal step : ell deformed Matsubara frequency summation 8} and \eqref{cal step : ell deformed Matsubara frequency summation 9}, one finds
\begin{align}
    \frac{1}{\beta} \sum_n \log(\tilde{\omega}_n^2 + \epsilon_{\vec{k}}^2)
& =  \frac{1}{\beta} \log\left(1 - \ell e^{-\beta \epsilon_{\vec{k}}} \right) +\frac{1}{2} \epsilon_{\vec{k}} + \tilde{C}_{1}(\beta, \ell) - \frac{1}{2} \epsilon_{\vec{k}} + \frac{1}{\beta} \log\left(1 - \ell e^{\beta \epsilon_{\vec{k}}} \right) + \tilde{C}_{2}(\beta, \ell) \notag \\
& = \frac{1}{\beta} \log(\ell^2 +1  -2 \ell \cosh(\beta \epsilon_{\vec{k}} )) +  \tilde{C}(\beta, \ell),
\end{align}
where $\tilde{C}_1(\beta, \ell) + \tilde{C}_2(\beta, \ell) = \tilde{C}(\beta, \ell)$.
The integration constants $\tilde{C}_1(\beta,\ell)$, $\tilde{C}_2(\beta,\ell)$ and $\tilde{C}(\beta,\ell)$ are independent of $\epsilon_{\vec{k}}$.

\section{ Derivation of the Jacobian and Fokker-Planck equations}
\label{Appendix C : Derivation of the Jacobian and Fokker-Planck equations}
\paragraph{Calculation of Jacobian factor}
When we construct a partition function and change of the variable, we need the Jacobian factor.
We define
\begin{equation}
\label{kernel A}
A(x,t;x',t')
=\delta^{(d)}(x-x')\partial_t\delta(t-t'),
\qquad
A^{-1}(x,t;x',t')=\delta^{(d)}(x-x')\theta(t-t'),
\end{equation}
where A is a kernel satisfying, $AA^{-1}=1$.
\begin{align}
&\int dx'\int dt'A(x,t;x',t')A^{-1}(x',t';x'',t'')
=\int dx'\int dt' \delta^{(d)}(x-x') \delta^{(d)}(x'-x'')\partial_t\delta(t-t')\theta(t'-t'') \notag \\
&= \delta^{(d)}(x-x'') \big[ -\int dt'  \partial_{t'}\delta(t-t')\theta(t'-t'') \big]= \delta^{(d)}(x-x'') \big[ \int dt' \delta(t-t')\partial_{t'}\theta(t'-t'') \big] \notag \\
&=\delta^{(d)}(x-x'')\delta(t-t'').
\end{align}
Here, $AA^{-1}=1$ are written as 1 for shorthand notation, but the identity operator is explicitly represented by the expression above.
Now consider the Jacobian factor
\begin{align}
\label{cal Jacobian factor}
\left|
\frac{\delta \eta(x,t)}{\delta \phi(x',t')}
\right|
& = \left| \frac{\delta}{\delta\phi(x',t')} \left[\dot{\phi}(x,t) + \frac{1}{2}\frac{\delta S_{cl}}{\delta\phi(x,t)}\right] \right|.
\end{align}
To evaluate the first term in \eqref{cal Jacobian factor}
\begin{align}
\frac{\delta\dot{\phi}(x,t)}{\delta\phi(x',t')} & =
\delta^{(d)}(x-x')\partial_t\delta(t-t'), \notag \\
& =  \int d^d x'' d t'' \delta^{(d)} (x-x'') \delta(t-t'') \delta^{(d)}(x''-x') \partial_{t''}\delta (t''-t').
\end{align}
and evaluate second term in \eqref{cal Jacobian factor}


\begin{align}
\frac{1}{2}
\frac{\delta}{\delta\phi(x',t')}
\left[\frac{\delta S_{cl}}{\delta\phi(x,t)}\right]
&=
\int d^d x'' d t'' \frac{1}{2}\frac{\delta^2 S_{cl}}{\delta\phi(x,t)\delta\phi(x'',t'')} \frac{\delta \phi(x'',t'')}{\delta \phi (x',t')}, \notag\\ 
&=\int d^d x'' d t'' \frac{1}{2}\frac{\delta^2 S_{cl}}{\delta\phi(x,t)\delta\phi(x'',t'')} \delta^{(d)}(x''-x')\delta(t''-t') .
\end{align}
The result \eqref{cal Jacobian factor} is 
\begin{align}
\left|
\frac{\delta \eta(x,t)}{\delta \phi(x',t')}
\right|
&=
\left|
\int d^d \tilde{x} d\tilde{t} \delta^{(d)}(\tilde{x}-x') \partial_{\tilde{t}}\delta(\tilde{t}-t') \right. \notag
\\
&\quad \left.
\times
\left[ \delta^{(d)}(x-\tilde{x})\delta(t-\tilde{t})
+ \frac12 \int d^d x'' dt'' \frac{\delta^2 S_{cl}}{\delta\phi(x,t)\delta\phi(x'',t'')} \delta^{(d)}(x''-\tilde{x})
\theta(t''-\tilde{t})\right] 
\right|.
\end{align}
To simplify this, we define $B(x,t;x',t')$,
\begin{align}
A(x,t;x',t')=& \int d^d \tilde{x} d\tilde{t} \delta^{(d)}(\tilde{x}-x') \partial_{\tilde{t}}\delta(\tilde{t}-t')\delta^{(d)}(x-\tilde{x})\delta(t-\tilde{t}) = \delta^{(d)}(x-x')\partial_t\delta(t-t'), \notag \\
B(x,t;x',t') =&  \int d^d \tilde{x} d\tilde{t} \delta^{(d)}(\tilde{x}-x') \partial_{\tilde{t}}\delta(\tilde{t}-t') \frac12 \int d^d x'' dt'' \frac{\delta^2 S_{cl}}{\delta\phi(x,t)\delta\phi(x'',t'')} \delta^{(d)}(x''-\tilde{x}) \theta(t''-\tilde{t}) \notag \\
& =\frac{1}{2}\frac{\delta^2 S_{cl}}{\delta\phi(x,t)\delta\phi(x',t')}.
\end{align}
The Jacobian factor \eqref{cal Jacobian factor} is decomposed into two determinants,
\begin{align}
\left|\frac{\delta\eta}{\delta\phi}\right|
&=\det(A+B) =\det(A)\det(1+A^{-1}B).
\end{align}
Here, $A=\delta^{(d)}(x-x') \partial_{t}\delta(t-t')$ is field independent, $\det A$ is a constant and can be absorbed into the overall normalization \cite{Moshe:2003xn}.
Hence only $\det(1+A^{-1}B)$ contributes.
Therefore,
\begin{equation}
\log\left|\frac{\delta\eta}{\delta\phi}\right|=\Tr\log(1+A^{-1}B).
\end{equation}
Expanding Taylor series,
\begin{equation}
\Tr\log(1+K)=\Tr K-\frac{1}{2}\Tr K^2+\frac{1}{3}\Tr K^3-\cdots,
\end{equation}
where $K(x,t;x'',t'') \equiv (A^{-1}B)(x,t;x'',t'').$
We will look at the simple case of $\Tr K$ to see how it works.
\begin{align*}
K(x,t;x'',t'')&= \int d^d x' \int dt' A^{-1}(x,t;,x',t')B(x',t;x'',t''), \\
&= \int d^d x' \int dt' \delta^{(d)}(x-x')\theta(t-t')\frac{1}{2}\frac{\delta^2 S_{cl}}{\delta\phi(x',t')\delta\phi(x'',t'')},\\
&=\frac{1}{2}\int dt' \theta(t-t')\frac{\delta^2 S_{cl}}{\delta\phi(x,t')\delta\phi(x'',t'')}.
\end{align*}
For $n\ge 2$, $\mathrm{Tr}K^n$ contains a factor, simply consider n=1 to 3 case. \\
For n=1 case.
\begin{align}
\mathrm{Tr}K & = \int d^d x dt  K(x,t;x,t), \notag \\
& = \int d^d x dt \int d^d x'' dt'' \frac{1}{2} \int dt' \theta(t-t')\frac{\delta^2 S_{cl}}{\delta\phi(x,t')\delta\phi(x'',t'')} \delta^{(d)}(x''-x)\delta(t''-t), \notag \\
& = \frac12 \int d^d x dt \int dt' \theta(t-t') \frac{\delta^2 S_{cl}}{\delta\phi(x,t')\delta\phi(x,t)}.
\end{align}
In our calculations, since the variation of $S_{cl}$ is given at the same location x, if we assume that $S_{cl}$ is given locally, $S_{cl}$ can be thought of in the following form
\begin{align}
     S_{cl}(t) & =\int d^dy  g(t) \phi^{n}(y,t), \\
     \frac{\delta S_{cl}(t)}{\delta \phi(x,t)} & = \int d^dy  ng(t) \phi^{n-1}(y,t) \delta^{(d)}(y-x) = ng(t)\phi^{n-1}(x,t), \notag \\ 
     \label{same time local derivative of S}
     \frac{\delta^{2} S_{cl}(t)}{\delta \phi(x,t)\delta \phi(x,t)} & = \frac{\delta}{\delta \phi(x,t)} \big[ ng(t)\phi^{n-1}(x,t) \big] = n(n-1)g(t)\phi^{n-2}(x,t)\delta^{(d)}(0),
\end{align}
Similarly,
\begin{equation}
\label{diff time local derivative of S}
    \frac{\delta^{2} S_{cl}(t)}{\delta \phi(x,t)\delta \phi(x,t')}  = \frac{\delta}{\delta \phi(x,t')} \big[ ng(t)\phi^{n-1}(x,t) \big] = n(n-1)g(t)\phi^{n-2}(x,t)\delta^{(d)}(0)\delta(t-t'). 
\end{equation}
Comparing \eqref{same time local derivative of S} and \eqref{diff time local derivative of S}, we get
\begin{equation}
     \frac{\delta^{2} S_{cl}(t)}{\delta \phi(x,t)\delta \phi(x,t)} \delta(t-t') =  \frac{\delta^{2} S_{cl}(t)}{\delta \phi(x,t)\delta \phi(x,t')}.
\end{equation}
Finally, we get
\begin{align}
    \mathrm{Tr}K & = \frac12 \int d^d x dt \int dt' \theta(t-t') \frac{\delta^2 S_{cl}(t)}{\delta\phi(x,t')\delta\phi(x,t)} \notag \\
    & = \frac12 \int d^d x dt \int dt' \theta(t-t')\delta(t-t')\frac{\delta^{2} S_{cl}(t)}{\delta \phi(x,t)\delta \phi(x,t)} \notag \\
    \label{calculated Trace K}
    & = \frac12 \theta(0) \int d^d x dt  \frac{\delta^{2} S_{cl}(t)}{\delta \phi(x,t)\delta \phi(x,t)}.
\end{align}
For n=2 and n=3, consider the products of $\theta$ functions
\begin{align}
\label{zero measure part cal in general step}
\mathrm{Tr}K^2 &= \int d^d x dt \int d^d x' dt' K(x,t;x',t') K(x',t';x,t), \\
& \sim \int dt \int dt' \theta(t-t')\theta(t'-t)=0, \\
\mathrm{Tr}K^3 & = \int d^d x dt \int d^d x' dt' \int d^d x'' dt''  K(x,t;x',t') K(x',t';x'',t'')K(x'',t'';x,t), \\
& \sim \int dt \int dt' \int dt'' \theta(t-t')\theta(t'-t'')\theta(t''-t) =0, 
\end{align}
which would require $t>t_1>\cdots>t_n>t$. 
To simplify the situation, we focus on two conditions $t>t_1>\cdots>t_n, t_n>t$.
These two requirements are clearly incompatible. This inconsistency originates from taking the trace, which enforces a cyclic time ordering. Indeed, $\mathrm{Tr}\,K^n$ contains the chain of theta functions $\theta(t-t_1)\theta(t_1-t_2)\cdots\theta(t_{n-1}-t)$.
Namely, at least one of the theta factors must vanish at $\mathrm{Tr}\,K^n=0$ for $n\ge2$. But the last condition left to resolve this contradiction is that all times can survive if they are equal time.
It is important to note that when $t=t'$, a factor of $\theta(0)$ arises from the trace. Naively, one might expect that $\theta(0)$ appears $n$ times in $\mathrm{Tr} K^n$, give a finite contribution.
However, for $n=2$, the integration measure is two-dimensional, $dt,dt'$, while the contribution of $\theta(t-t')\theta(t'-t)$ is restricted to $t=t'$, vanish. Similarly, for $n=3$, the integration measure is three-dimensional, $dt,dt',dt''$, but the  contribution of step functions is restricted to $t=t'=t''$, which again has measure zero, and thus the integral vanishes.
Thus, taking $\theta(0)=1/2$,
\begin{equation}
\Tr K =\frac14  \int d^d x dt \frac{\delta^{2} S_{cl}(t)}{\delta \phi(x,t)\delta \phi(x,t)},
\end{equation}
\begin{equation}
\Tr K=-\bigg[-\frac{1}{4} \int d^dxdt\frac{\delta^2 S_{cl}(t)}{\delta\phi^2(x,t)}\bigg].
\end{equation}
Trace notation is given by
\begin{align*}
& \Tr(A) = \Sigma_n \langle n |A | n \rangle, \qquad 1 = \int dx |x\rangle \langle x|, \\
& \Tr(A) = \Sigma_n \langle n |A1| n \rangle  = \Sigma_n \int dx \langle n |A|x \rangle \langle x | n \rangle = \int dx \langle x |A | x\rangle.
\end{align*}
We compute the cross term $\frac{\delta S_{cl}}{\delta\phi}\partial_t\phi(x,t)$,
\begin{equation}
\label{exist cross term Scl}
\int_0^t dt'\int d^dx\partial_t\phi(x,t') \frac{\delta S_{cl}}{\delta\phi(x,t')}.
\end{equation}
Substituting \eqref{O derivative of Scl} into \eqref{exist cross term Scl},
\begin{equation}
\label{O derivative of Scl}
\frac{dS_{cl}[\phi]}{dt}=\int d^dx\frac{\delta S_{cl}}{\delta\phi}\partial_t\phi(x,t).
\end{equation}
We obtain
\begin{equation}
\int_0^t dt'\int d^dx\partial_t\phi(x,t')\frac{\delta S_{cl}}{\delta\phi(x,t')}=\int_0^t dt'\frac{dS_{cl}[\phi(t')]}{dt'}=S_{cl}[\phi(t)]-S_{cl}[\phi(0)].
\end{equation}
The partition function is given by
\begin{equation}
Z=\int D\phi(x,0)P[\phi,0]\prod_{0<t'<t}D\phi(x,t')
\exp\left[-\frac{1}{\hbar}\int_0^t dt'\int d^dx\mathcal{L}_{FP}[\phi(x,t'),t']\right]
D\phi(x,t)P[\phi,t].
\end{equation}
The Fokker-Planck Lagrangian,  $2\lambda=\hbar$,
\begin{align}
\mathcal{L}_{FP}
=
\frac{1}{2}\left(\frac{\partial\phi}{\partial t}\right)^2
+
\frac{1}{8}
\left(
\frac{\delta S_{cl}}{\delta\phi}
\right)^2
-
\frac{\hbar}{4}
\frac{\delta^2 S_{cl}}{\delta\phi^2}.
\end{align}
\paragraph{Derivation of the Fokker-Planck equation}
We start from the definition of the probability functional.
Here, we keep $\lambda$ and start the expansion. 
However, as can be seen in the main text, the wave function in the equilibrium is ultimately $~\frac{1}{\hbar} S_{cl}$, so we need to write $2\lambda = \hbar$ to match it. 
Also, since the above relationship is given linearly by the Einstein relation, there is no problem keeping $\lambda$ and converting to $\hbar$ at the end. The probability functional is
\begin{equation}
P[\phi,t]=\int D\eta P[\eta]\delta[\phi-\phi_\eta(t)]
=\left\langle \delta[\phi-\phi_\eta(t)] \right\rangle_\eta.
\end{equation}
Taking the (stochastic) time derivative,
\begin{align}
\partial_t P[\phi,t]=\left\langle \partial_t \delta[\phi-\phi_\eta(t)] \right\rangle_\eta.
\end{align}
Using the functional chain rule,
\begin{align}
\partial_t \delta[\phi-\phi_\eta(t)]
&=-\int d^dx \dot{\phi}_\eta(x,t)\frac{\delta}{\delta\phi(x)}
\delta[\phi-\phi_\eta(t)].
\end{align}
Taking the expectation value with respect to the noise distribution $P[\eta]$ both sides, we obtain
\begin{equation}
\label{Probability evolution equation}
\partial_t P[\phi,t]=-\int d^d x\left\langle\dot{\phi}_\eta(x,t)\frac{\delta}{\delta\phi(x)}\delta[\phi-\phi_\eta(t)]\right\rangle_\eta.
\end{equation}
Since the functional derivative $\delta/ \delta\phi(x)$ acts only on the dummy field configuration $\phi(x)$, i.e. $\phi(x)$ is some value, and $\dot{\phi}_\eta(x,t)$ is independent of $\phi(x)$, we rewrite
\begin{equation}
\partial_t P[\phi,t]=-\int d^dx\frac{\delta}{\delta\phi(x)}\left\langle
\dot{\phi}_\eta(x,t)\delta[\phi-\phi_\eta(t)]\right\rangle_\eta.
\end{equation}
Now substitute the Langevin equation, 
\begin{equation}
\dot{\phi}_\eta(x,t)
=-\frac{1}{2}\frac{\delta S_{cl}}{\delta\phi_{\eta}(x,t)}+\eta(x,t),
\end{equation}
where $k=1/2$.
Then,
\begin{align}
\label{cal step derivation of the FP eq 1}
\left\langle\dot{\phi}_\eta(x,t)\delta[\phi-\phi_\eta(t)]\right\rangle_\eta
&=-\frac{1}{2}\left\langle\frac{\delta S_{cl}}{\delta\phi(x,t)}\delta[\phi-\phi_\eta(t)]\right\rangle_\eta+\left\langle\eta(x,t)\delta[\phi-\phi_\eta(t)]\right\rangle_\eta.
\end{align}
To evaluate the first term in \eqref{cal step derivation of the FP eq 1},
\begin{align}
-\frac{1}{2}\left\langle\frac{\delta S_{cl}}{\delta\phi(x,t)}\delta[\phi-\phi_\eta(t)]\right\rangle_\eta
&=-\frac{1}{2}\frac{\delta S_{cl}}{\delta\phi_{\eta}(x,t)}\left\langle\delta[\phi-\phi_\eta(t)]\right\rangle_\eta,
\notag\\
&=-\frac{1}{2}\frac{\delta S_{cl}}{\delta\phi_{\eta}(x,t)}P[\phi,t].
\end{align}
Now consider the second term in \eqref{cal step derivation of the FP eq 1}.
Using the Gaussian noise distribution(Z is inverse of normalization factor N, $Z=N^{-1}$),
\begin{equation}
P[\eta]=\frac{1}{Z}\exp\left[-\frac{1}{4\lambda}\int d^dxdt\,\eta(x,t)^2
\right].
\end{equation}
We get
\begin{equation}
\frac{\delta P[\eta]}{\delta\eta(x,t)}=-\frac{1}{2\lambda}\eta(x,t)P[\eta].
\end{equation}
Equivalently,
\begin{equation}
\eta(x,t)P[\eta]=-2\lambda\frac{\delta P[\eta]}{\delta\eta(x,t)}.
\end{equation}
Therefore, the second term in \eqref{cal step derivation of the FP eq 1} is
\begin{align}
\label{noise integral cal1}
\left\langle\eta(x,t)\delta[\phi-\phi_\eta(t)]\right\rangle_\eta
&=\int D\eta P[\eta]\eta(x,t)\delta[\phi-\phi_\eta(t)]
\notag\\
&=-2\lambda\int D\eta\frac{\delta P[\eta]}{\delta\eta(x,t)}\delta[\phi-\phi_\eta(t)].
\end{align}
Using integration by parts, the boundary term vanishes,
\begin{equation}
\int D\eta\frac{\delta}{\delta\eta(x,t)}\left(P[\eta]\delta[\phi-\phi_\eta(t)]
\right)=0.
\end{equation}
We obtain
\begin{equation}
\label{noise integral cal2}
\int D\eta\frac{\delta P[\eta]}{\delta\eta(x,t)}\delta[\phi-\phi_\eta(t)]
= -\int D\eta P[\eta]\frac{\delta}{\delta\eta(x,t)}\delta[\phi-\phi_\eta(t)].
\end{equation}
Substituting \eqref{noise integral cal2} to \eqref{noise integral cal1}, we get
\begin{equation}
\left\langle \eta(x,t)\delta[\phi-\phi_\eta(t)] \right\rangle_\eta
=2\lambda \left\langle \frac{\delta}{\delta\eta(x,t)} \delta[\phi-\phi_\eta(t)] \right\rangle_\eta.
\end{equation}
Using chain rule,
\begin{equation}
\frac{\delta}{\delta\eta(x,t)}\delta[\phi-\phi_\eta(t)]
=-\int d^dy\frac{\delta\phi_\eta(y,t)}{\delta\eta(x,t)}\frac{\delta}{\delta\phi(y)}\delta[\phi-\phi_\eta(t)].
\end{equation}
From the Langevin equation,
\begin{equation}
\phi_\eta(y,t)-\phi_\eta(y,0)
=\int_0^t dt'\left[-\frac{1}{2}\frac{\delta S_{cl}}{\delta\phi_{\eta}(y,t')} + \eta(y,t')\right].
\end{equation}
Similarly, we take the functional derivative
\begin{align}
\frac{\delta\phi_\eta(y,t)}{\delta\eta(x,\tilde{t})} & \equiv F(y,t;x,\tilde{t})
=\frac{\delta}{\delta\eta(x,\tilde{t})}\int_0^t dt'\left[-\frac{1}{2}\frac{\delta S_{cl}}{\delta\phi_{\eta}(y,t')} + \eta(y,t')\right]\notag\\
&=\delta^{(d)}(y-x)\theta(t'-\tilde{t})|_{0}^{t} -\frac12 \int_{0}^{\infty} dt' \theta(t-t') \int_{0}^{t} dt'' \int d^dz \frac{\delta^{2} S_{cl}}{\delta \phi_{\eta}(y,t')\delta \phi_{\eta}(z,t'')} \frac{\delta \phi_{\eta}(z,t'')}{\delta \eta (x,\tilde{t})}.
\end{align}
This is often used in the typical iterative form $F=A+KA$, so we know that $F = A+KA+KKA+$ $\cdots $ for $|K|<<1$:
\begin{align*}
    F^{0}(y,t;x,\tilde{t}) & = \delta^{(d)}(y-x)\theta(t'-\tilde{t})|_{0}^{t} = \delta^{(d)}(y-x)\theta(t-\tilde{t}) \text{ for } 0<\tilde{t}<t'<t, \\
     F^{1}(y,t;x,\tilde{t}) & = -\frac12 \int_{0}^{\infty} dt' \theta(t-t') \int_{0}^{t} dt'' \int d^dz \frac{\delta^{2} S_{cl}}{\delta \phi_{\eta}(y,t')\delta \phi_{\eta}(z,t'')} F^{0}(z,t'';x,\tilde{t}) \notag \\
     & = -\frac12 \int_{0}^{\infty} dt' \theta(t-t') \int_{0}^{\infty} dt'' \theta(t-t'') \int d^dz \frac{\delta^{2} S_{cl}}{\delta \phi_{\eta}(y,t')\delta \phi_{\eta}(z,t'')}  \delta^{(d)}(z-x)\theta(t''-\tilde{t}).
\end{align*}
Also, we apply the equal time relation $\tilde{t} = t$,
\begin{align}
    F^{0}(y,t;x,\tilde{t}) & = \delta^{(d)}(y-x)\theta(0), \\
     F^{1}(y,t;x,\tilde{t}) & = -\frac12 \int_{0}^{\infty} dt' \theta(t-t') \int_{0}^{\infty} dt'' \theta(t-t'') \int d^dz \frac{\delta^{2} S_{cl}}{\delta \phi_{\eta}(y,t')\delta \phi_{\eta}(z,t'')}  \delta^{(d)}(z-x)\theta(t''-t).
\end{align}
As explained before \eqref{zero measure part cal in general step}
\begin{equation*}
    \sim \int dt \int dt' \theta(t-t')\theta(t'-t)=0.
\end{equation*}
Only the $F^{0}$ term survives and we get
\begin{equation}
    F(y,t;x,t)=  \delta^{(d)}(y-x)\theta(0).
\end{equation}
Using $\theta(0)=1/2$,
\begin{equation}
\frac{\delta\phi_\eta(y,t)}{\delta\eta(x,t)}= F(y,t;x,t)=\frac{1}{2}\delta^{(d)}(y-x).
\end{equation}
Hence,
\begin{equation}
\frac{\delta}{\delta\eta(x,t)}\delta[\phi-\phi_\eta(t)]=-\frac{1}{2}\frac{\delta}{\delta\phi(x,t)}\delta[\phi-\phi_\eta(t)].
\end{equation}
The result of the second term in \eqref{cal step derivation of the FP eq 1} is
\begin{align}
\left\langle\eta(x,t)\delta[\phi-\phi_\eta(t)]\right\rangle_\eta
&=2\lambda\left\langle-\frac{1}{2}\frac{\delta}{\delta\phi(x,t)}\delta[\phi-\phi_\eta(t)]\right\rangle_\eta\notag\\
&= -\lambda\frac{\delta}{\delta\phi(x,t)}\left\langle\delta[\phi-\phi_\eta(t)]\right\rangle_\eta = -\lambda\frac{\delta}{\delta\phi(x,t)}P[\phi,t].
\end{align}
The result of \eqref{cal step derivation of the FP eq 1} is given as
\begin{equation}
\left\langle\dot{\phi}_\eta(x,t)\delta[\phi-\phi_\eta(t)]\right\rangle_\eta
=-\frac{1}{2}\frac{\delta S_{cl}}{\delta\phi(x,t)}P[\phi,t] -\lambda\frac{\delta}{\delta\phi(x,t)}P[\phi,t].
\end{equation}
Substituting into the Probability evolution equation \eqref{Probability evolution equation},
\begin{align}
\partial_t P[\phi,t]
&=-\int d^dx\frac{\delta}{\delta\phi(x,t)}\left[-\frac{1}{2}\frac{\delta S_{cl}}{\delta\phi(x,t)}P[\phi,t]-\lambda\frac{\delta}{\delta\phi(x,t)}P[\phi,t]\right].
\end{align}
Finally, we get the Fokker-Planck equation via stochastic process, keep $\lambda$,
\begin{equation}
\label{Fokker-Planck equation}
\partial_t P[\phi,t]=\frac{1}{2} \int d^dx\frac{\delta}{\delta\phi(x,t)}\left[\frac{\delta S_{cl}}{\delta\phi(x,t)} + 2\lambda\frac{\delta}{\delta\phi(x,t)} \right]P[\phi,t].
\end{equation}
Here, we conveniently define and use wave functions related to Schr\"odinger-type equations.
\begin{equation}
P[\phi]=e^{-\frac{1}{4\lambda}S_{cl}[\phi]}\psi_s[\phi].
\end{equation}
We now expand the following functional derivative
\begin{equation}
\frac{\delta}{\delta\phi} \left(\frac{\delta S_{cl}}{\delta\phi}P+
2\lambda\frac{\delta P}{\delta\phi}\right).
\end{equation}
Take another one
\begin{align}
\frac{\delta}{\delta\phi}\left(\frac{\delta S_{cl}}{\delta\phi}P+2\lambda\frac{\delta P}{\delta\phi}\right)
&=\frac{\delta^2 S_{cl}}{\delta\phi^2}P+\frac{\delta S_{cl}}{\delta\phi}\frac{\delta P}{\delta\phi}+2\lambda\frac{\delta^2 P}{\delta\phi^2}.
\end{align}
Now we substitute
$P=e^{-\frac{1}{4\lambda}S_{cl}}\psi_s$
and compute each term explicitly.
First, the functional derivative of $P$ is
\begin{align}
\frac{\delta P}{\delta\phi}
&=\frac{\delta}{\delta\phi}\left(e^{-\frac{1}{4\lambda}S_{cl}}\psi_s
\right), \notag \\
&=\left(-\frac{1}{4\lambda}\frac{\delta S_{cl}}{\delta\phi}\psi_s + \frac{\delta\psi_s}{\delta\phi}
\right) e^{-\frac{1}{4\lambda}S_{cl}}.
\end{align}
Next, taking one more functional derivative, we obtain
\begin{align}
\frac{\delta^2 P}{\delta\phi^2}
&=\frac{\delta}{\delta\phi}\left[\left(-\frac{1}{4\lambda}\frac{\delta S_{cl}}{\delta\phi}\psi_s + \frac{\delta\psi_s}{\delta\phi}\right)e^{-\frac{1}{4\lambda}S_{cl}} \right].
\end{align}
Collecting the part of calculation,
\begin{align}
\frac{\delta^2 P}{\delta\phi^2}
&=\left[ -\frac{1}{4\lambda}\frac{\delta^2 S_{cl}}{\delta\phi^2}\psi_s
-\frac{1}{4\lambda}\frac{\delta S_{cl}}{\delta\phi}\frac{\delta\psi_s}{\delta\phi} + \frac{\delta^2\psi_s}{\delta\phi^2}\right]
e^{-\frac{1}{4\lambda}S_{cl}} + \left( -\frac{1}{4\lambda}\frac{\delta S_{cl}}{\delta\phi}\psi_s + \frac{\delta\psi_s}{\delta\phi} \right)
\frac{\delta}{\delta\phi} \left( e^{-\frac{1}{4\lambda}S_{cl}} \right).
\end{align}
The remaining derivative term is
\begin{align}
\frac{\delta}{\delta\phi}\left(e^{-\frac{1}{4\lambda}S_{cl}}
\right)= - \frac{1}{4\lambda}\frac{\delta S_{cl}}{\delta\phi}
e^{-\frac{1}{4\lambda}S_{cl}}.
\end{align}
Substituting above relation, we get
\begin{align}
\frac{\delta^2 P}{\delta\phi^2}
&=\Bigg[ -\frac{1}{4\lambda}\frac{\delta^2 S_{cl}}{\delta\phi^2}\psi_s
-\frac{1}{4\lambda}\frac{\delta S_{cl}}{\delta\phi}\frac{\delta\psi_s}{\delta\phi} +\frac{\delta^2\psi_s}{\delta\phi^2}
+ \frac{1}{16\lambda^2} \left(\frac{\delta S_{cl}}{\delta\phi}\right)^2\psi_s
-\frac{1}{4\lambda}\frac{\delta S_{cl}}{\delta\phi}\frac{\delta\psi_s}{\delta\phi} \Bigg] e^{-\frac{1}{4\lambda}S_{cl}} \notag \\
&=\Bigg[
-\frac{1}{4\lambda}\frac{\delta^2 S_{cl}}{\delta\phi^2}\psi_s
+\frac{1}{16\lambda^2} \left(\frac{\delta S_{cl}}{\delta\phi}\right)^2\psi_s
+\frac{\delta^2\psi_s}{\delta\phi^2} -\frac{1}{2\lambda}\frac{\delta S_{cl}}{\delta\phi}\frac{\delta\psi_s}{\delta\phi} \Bigg] e^{-\frac{1}{4\lambda}S_{cl}}.
\end{align}
Now we insert $\frac{\delta P}{\delta\phi}$ and $\frac{\delta^2 P}{\delta\phi^2}$ into the original expansion \eqref{Fokker-Planck equation}.
Then the cross terms proportional to
$\frac{\delta S_{cl}}{\delta\phi}\frac{\delta\psi_s}{\delta\phi}$
cancel, and we obtain
\begin{align}
\label{Fokker-Planck equation of full cal1}
\frac{\delta}{\delta\phi} \left( \frac{\delta S_{cl}}{\delta\phi}p +
2\lambda\frac{\delta p}{\delta\phi} \right)
&= e^{-\frac{1}{4\lambda}S_{cl}}
\Bigg[ \frac{1}{2}\frac{\delta^2 S_{cl}}{\delta\phi^2}\psi_s
-\frac{1}{8\lambda} \left(\frac{\delta S_{cl}}{\delta\phi}\right)^2\psi_s
+2\lambda\frac{\delta^2\psi_s}{\delta\phi^2} \Bigg].
\end{align}
Requiring $\psi_S$ to satisfy a Schr\"odinger-type evolution equation, one obtains
\begin{equation}
 \frac{\partial\psi_S(\phi,t)}{\partial t} = - \frac{1}{\hbar}\int d^d x \mathcal{H}_{\mathrm{FP}}(\frac{\delta}{\delta \phi},\phi) \psi_S(\phi,t), \ \ (2\lambda = \hbar)
\end{equation}
where $\mathcal{H}_{FP}$ is Fokker-Planck Hamiltonian, which is given by
\begin{equation}
\mathcal{H}_{FP}
= -\frac{\hbar^2}{2}\frac{\delta^2}{\delta\phi^2} + \frac{1}{8}
 \left(\frac{\delta S_{cl}}{\delta\phi}\right)^2 - \frac{\hbar}{4}\frac{\delta^2 S_{cl}}{\delta\phi^2}.
\end{equation}
Finally, this operator can be factorized
\begin{align}
\mathcal{H}_{FP}
& = \frac{\hbar^2}{2}\left( -\frac{\delta}{\delta\phi} + \frac{1}{2 \hbar}\frac{\delta S_{cl}}{\delta\phi} \right) \left( \frac{\delta}{\delta\phi}
+\frac{1}{2\hbar}\frac{\delta S_{cl}}{\delta\phi} \right).
\end{align}
Note that $S_{cl}=S_{cl}[\phi(t);t], \frac{d S_{\mathrm{cl}}[\phi]}{d t}=\int d^d x \frac{\delta S_{\mathrm{cl}}}{\delta \phi(x,t)} \partial_t \phi(x,t) +\frac{\partial S_{cl}}{\partial t}$, but $S_{cl}=S_{cl}[\phi(t)]$, $\partial_t S_{cl}=0$.
The derivation above corresponds to the case $S_{cl}=S_{cl}[\phi(t)].$
For an explicitly time dependent action $S_{cl}[\phi(t);t]$, the Schr\"odinger type evolution acquires the additional term $-\frac{1}{2}\partial_t S_{cl}$ in the effective Fokker-Planck Hamiltonian, i.e. $\partial_t \psi_s = -\hbar^{-1}\int d^dx \left[   \mathrm{H}_{FP}-\frac{1}{2} \partial_t S_{cl}   \right]$.
\end{appendices}


\begin{thebibliography}{99}

\bibitem{Damgaard:1987rr}
P.~H.~Damgaard and H.~Huffel,
Phys. Rept. \textbf{152}, 227 (1987)
doi:10.1016/0370-1573(87)90144-X

\bibitem{Dijkgraaf:2009gr}
R.~Dijkgraaf, D.~Orlando and S.~Reffert,
Nucl. Phys. B \textbf{824}, 365-386 (2010)
doi:10.1016/j.nuclphysb.2009.07.018
[arXiv:0903.0732 [hep-th]].


\bibitem{Mansi:2009mz}
D.~S.~Mansi, A.~Mauri and A.~C.~Petkou,
Phys. Lett. B \textbf{685}, 215-221 (2010)
doi:10.1016/j.physletb.2010.01.033
[arXiv:0912.2105 [hep-th]].





\bibitem{1}
O. Aharony, S. S. Gubser, J. Maldacena, H. Ooguri, and Y. Oz, Large N Field Theories, String Theory and Gravity, Phys. Rept. 323, 183–386 (2000), [arXiv:hep-th/9905111]
\bibitem{2}
Udo Seifert, Entropy Production along a Stochastic Trajectory and an Integral Fluctuation Theorem, Phys. Rev. Lett. 95, 040602, [arXiv:cond-mat/0503686]
\bibitem{3}
Jae-Hyuk Oh, Dileep P. Jatkar, Stochastic quantization and holographic Wilsonian renormalization group, [arXiv:hep-th/1209.2242]




\bibitem{5}
Michael E. Peskin and Daniel V. Schroeder, An Introduction to Quantum Field Theory, Addison-Wesley, 1995, p463.
\bibitem{6}
Steven S. Gubser, Indrajit Mitra, Double-trace operators and one-loop vacuum energy in AdS/CFT, Phys.Rev.D67, 064018 (2003), [arXiv:hep-th/0210093]
\bibitem{7}
Edward Witten, Multi-Trace Operators, Boundary Conditions, And AdS/CFT Correspondence, [arXiv:hep-th/0112258]
\bibitem{8}
Thomas Hartman, Leonardo Rastelli, Double-Trace Deformations, Mixed Boundary Conditions and Functional Determinants in AdS/CFT, JHEP0801:019,2008, [arXiv:hep-th/0602106]
\bibitem{9}
A. C. Petkou, M. B. Silva Neto, On the Free-Energy of Three-Dimensional CFTs and Polylogarithms, Phys.Lett. B456 147-154 (1999), [arXiv:hep-th/9812166]
\bibitem{10}
Gerald V. Dunne, Mithat Ünsal, Resurgence and Trans-series in Quantum Field Theory: The CP(N-1) Model, JHEP 11 (2012) 170, [arXiv:hep-th/1210.2423]
\bibitem{11}
K. Skenderis, Lecture notes on holographic renormalization, 
Class. Quant. Grav. 19 (2002) 5849–5876, [arXiv:hep-th/0209067].






\bibitem{4}
Andrey V. Chubukov, Subir Sachdev, Jinwu Ye, Theory of Two-Dimensional Quantum Heisenberg Antiferromagnets with a Nearly Critical Ground State
, Phys.Rev.B49, 11919 (1994), [arXiv:cond-mat/9304046]

\bibitem{Moshe:2003xn}
M.~Moshe and J.~Zinn-Justin,
Phys. Rept. \textbf{385}, 69-228 (2003)
doi:10.1016/S0370-1573(03)00263-1
[arXiv:hep-th/0306133 [hep-th]].





\bibitem{Klebanov:2002ja}
I.~R.~Klebanov and A.~M.~Polyakov,
Phys. Lett. B \textbf{550}, 213-219 (2002)
doi:10.1016/S0370-2693(02)02980-5
[arXiv:hep-th/0210114 [hep-th]].

\bibitem{Giombi:2009wh}
S.~Giombi and X.~Yin,
JHEP \textbf{09}, 115 (2010)
doi:10.1007/JHEP09(2010)115
[arXiv:0912.3462 [hep-th]].


\bibitem{Giombi:2013fka}
S.~Giombi and I.~R.~Klebanov,
JHEP \textbf{12}, 068 (2013)
doi:10.1007/JHEP12(2013)068
[arXiv:1308.2337 [hep-th]].

\bibitem{Giombi:2016ejx}
S.~Giombi,
doi:10.1142/9789813149441{\_}0003
[arXiv:1607.02967 [hep-th]].


\bibitem{Maldacena:2011jn}
J.~Maldacena and A.~Zhiboedov,
J. Phys. A \textbf{46}, 214011 (2013)
doi:10.1088/1751-8113/46/21/214011
[arXiv:1112.1016 [hep-th]].


\bibitem{Oh:2012bx}
J.~H.~Oh and D.~P.~Jatkar,
JHEP \textbf{11}, 144 (2012)
doi:10.1007/JHEP11(2012)144
[arXiv:1209.2242 [hep-th]].

\bibitem{Jatkar:2013uga}
D.~P.~Jatkar and J.~H.~Oh,
JHEP \textbf{10}, 170 (2013)
doi:10.1007/JHEP10(2013)170
[arXiv:1305.2008 [hep-th]].


\bibitem{Oh:2013tsa}
J.~H.~Oh,
Int. J. Mod. Phys. A \textbf{29}, 1450082 (2014)
doi:10.1142/S0217751X14500821
[arXiv:1310.0588 [hep-th]].


\bibitem{Oh:2015xva}
J.~H.~Oh,
Phys. Rev. D \textbf{94}, no.10, 105020 (2016)
doi:10.1103/PhysRevD.94.105020
[arXiv:1504.03046 [hep-th]].

\bibitem{Oh:2021bxx}
J.~H.~Oh,
J. Korean Phys. Soc. \textbf{79}, no.10, 903-917 (2021)
doi:10.1007/s40042-021-00320-x
[arXiv:2110.05013 [hep-th]].


\bibitem{Kim:2023bhp}
G.~Kim, J.~s.~Chae, W.~Shin and J.~H.~Oh,
Int. J. Mod. Phys. A \textbf{38}, no.21, 2350114 (2023)
doi:10.1142/S0217751X23501142
[arXiv:2305.18920 [hep-th]].



\bibitem{Lee:2023ynb}
J.~H.~Lee and J.~H.~Oh,
J. Korean Phys. Soc. \textbf{83}, no.9, 665-674 (2023)
doi:10.1007/s40042-023-00926-3
[arXiv:2308.10010 [hep-th]].

\bibitem{Chae:2024zyn}
J.~s.~Chae and J.~H.~Oh,
Eur. Phys. J. C \textbf{85}, no.3, 344 (2025)
doi:10.1140/epjc/s10052-025-14037-9
[arXiv:2402.03841 [hep-th]].


\end{thebibliography}
\end{document}